\magnification=\magstep1
\tolerance 500
\rightline{TAUP 2546-99}
\rightline{IASSNS-HEP-99-23}
\rightline{1 March, 1999}
\vskip  2 true cm
\centerline{\bf Representation of the Resonance}
\centerline{\bf of a } 
\centerline{\bf Relativistic Quantum Field Theoretical Lee-Friedrichs 
Model}
\centerline{\bf in}
\centerline{\bf Lax-Phillips Scattering Theory}
\smallskip
\centerline{Y. Strauss$^a$ and L.P. Horwitz$^b$\footnote{*}{On 
sabbatical leave from School of Physics and Astronomy, Raymond and 
Beverly Sackler Faculty of Exact Sciences,  Tel Aviv University,
 Ramat Aviv, Israel 69978, and  Department of
Physics, Bar Ilan University, Ramat Gan 52900, Israel.}}
\centerline{{}$^a$School of Physics and Astronomy}
\centerline{Raymond and Beverly Sackler Faculty of Exact Sciences}
\centerline{Tel Aviv University, Ramat Aviv 69978, Israel}
\centerline{{}$^b$School of Natural Sciences}
\centerline{Institute for Advanced Study, Princeton, NJ 08540}
\bigskip
{\it Abstract:\/} The quantum mechanical description of the evolution
of an unstable system defined initially as a state in a Hilbert space
at a given time does not provide a semigroup (exponential) decay law.
The Wigner-Weisskopf survival amplitude, describing reversible
quantum transitions, may be dominated by exponential type decay in
pole approximation at times not too short or too  long, but, in the
two channel case, for example, the pole residues are not orthogonal,
and the evolution does not correspond to a semigroup (experiments on
the decay of the neutral $K$-meson system strongly support the
semigroup evolution postulated by Lee, Oehme and Yang, and Yang and
Wu).  The scattering theory of Lax and Phillips, originally
developed for classical wave equations, has been recently extended to
the description of the evolution of resonant states in the framework
of quantum theory.  The resulting evolution law of the unstable system
is that of a semigroup, and the resonant state is a well-defined
function in the Lax-Phillips Hilbert space.  In this paper we apply
this theory to a relativistically covariant quantum field theoretical
form of the (soluble) Lee model.  We construct the translation
representations with the help of the wave operators, and show that the
resulting Lax-Phillips $S$-matrix is an inner function (the
Lax-Phillips theory is essentially a theory of translation invariant
subspaces).   In the special case that the $S$-matrix is a rational
inner function, we obtain the resonant state explicitly and analyze
its particle ($V,\,N,\, \theta$) content.  If there is an exponential
bound, the general case differs only by a so-called trivial inner
factor, which does not change the complex spectrum, but may affect the
wave function of the resonant state.  
\vfill
\eject
\noindent
{\bf 1. Introduction.}
\par The theory of Lax and Phillips$^1$ (1967), originally developed for the 
description of resonances in electromagnetic or acoustic scattering phenomena,
has been  used as a framework for the construction of a description of
 irreversible resonant phenomena in the
 quantum theory$^{2-5}$ (which we will refer to as the quantum Lax-Phillips 
theory). This leads to a time evolution of resonant states
which is of semigroup type, i.e., essentially exponential decay.  Semigroup
evolution is necessarily a property of irreversible processes$^6$. It 
appears experimentally that elementary particle decay, to a high 
degree of accuracy, follows a semigroup law, and hence such processes seem 
to be irreversible.    
\par The theory of Weisskopf and Wigner$^7$, which 
is based on the definition of the survival amplitude of the initial state
$\phi$ (associated with the unstable system) as the scalar 
product of that state
with the unitarily evolved state, 
$$  (\phi, e^{-iHt}\phi) \eqno(1.1)$$
cannot have exact exponential behavior$^8$. One can easily generalize this 
construction to the problem of more than one resonance$^{9,10}$. If $P$ is the 
projection operator into the  subspace of initial states
($N$-dimensional for $N$ resonances),
the reduced evolution operator is given by
$$Pe^{-iHt} P .\eqno(1.1')$$
 This operator cannot be an element of a semigroup.$^8$
\par  Experiments 
on the decay of 
the neutral $K$-meson system$^{11}$ show clearly that the 
phenomenological description of Lee, Oehme and Yang$^{12}$, and Wu and 
Yang$^{13}$, by means of a $2\times 2$ effective Hamiltonian which corresponds
to an exact semigroup evolution of the unstable system, provides a very
accurate description of the data. Is can be proved that the Wigner-Weisskopf
theory cannot provide a semigroup evolution law$^8$ and, thus, an
effective $2\times 2$ Hamiltonian cannot emerge in the framework of this
theory. Furthermore, it has been shown, using estimates based on
the quantum mechanical Lee-Friedrichs model$^{14}$, that the experimental
results appear to rule out the application of the Wigner-Weisskopf theory to
the decay of the neutral $K$-meson system. While the exponential decay
law can be exhibited explicitly in terms of a Gel'fand triple$^{15}$, the 
representation of the resonant state in this framework is in a Banach space
which does not coincide with the quantum mechanical Hilbert space, and
 does not have the properties of a Hilbert space, such as scalar products 
and the possibility of calculating expectation values.
One cannot compute physical properties other than the lifetimes in this way.
\par The quantum Lax-Phillips 
theory provides the possibility of constructing a fundamental theoretical 
description of the resonant system which has exact semigroup evolution,
and represents the resonance as a {\it state in the Hilbert space}. In the 
following, we describe briefly the structure of this theory, and give
some physical interpretation for the states of the Lax-Phillips Hilbert space. 
\par The Lax-Phillips theory is defined in a Hilbert space $\overline{\cal H}$
 of states which 
contains two distinguished subspaces, ${\cal D}_\pm$, called ``outgoing'' and 
``incoming''.  There is a unitary evolution law which we denote
 by $U(\tau)$, for which these subspaces are invariant in the following
sense:
$$\eqalign{ U(\tau) {\cal D}_+ &\subset {\cal D}_+ \qquad  \tau \geq 0 \cr
U(\tau) {\cal D}_- &\subset {\cal D}_- \qquad \tau \leq 0 \cr} \eqno(1.2)$$
\par The translates of ${\cal D}_\pm$ under $U(\tau)$ are dense, i.e.,
$$ {\overline {{\bigcup_\tau}\, U(\tau) {\cal D}_\pm}} =  {\overline {\cal H}}
  \eqno(1.3)$$
and the asymptotic property
$$ {\bigcap_\tau }\,U(\tau) {\cal D}_\pm = \emptyset \eqno(1.4)$$
is assumed. It follows from these properties that 
$$ Z(\tau) = P_+ U(\tau) P_-, \eqno(1.5)$$
where $P_\pm$ are projections into the subspaces orthogonal to ${\cal D}_\pm$,
is a strongly contractive semigroup$^1$, i.e.,
$$ Z(\tau_1) Z(\tau_2) = Z(\tau_1 + \tau_2) \eqno(1.6)$$
for $\tau_1,\, \tau_2$ positive, and $\Vert Z(\tau)\Vert \to 0$ for $\tau \to
0$. It follows from $(1.2)$ that $Z(\tau)$ takes the subspace
$\cal K$, the orthogonal complement of ${\cal D}_\pm$ in $\overline{\cal H}$
(associated with the resonances in the Lax-Phillips theory),
into itself$^1$, {\it i.e.},
$$ Z(\tau)= P_{\cal K} U(\tau) P_{\cal K}. \eqno(1.7)$$
The relation $(1.7)$ is of the same structure as $(1.1')$; there is,
 as we 
shall see in the following, an essential difference in the way that the
subspaces associated with resonances are defined. The argument that $(1.1')$
cannot form a semigroup is not valid$^3$ for $(1.7)$. 
\par There is a theorem of Sinai$^{16}$ which affirms that a Hilbert 
space with the properties that there are distinguished subspaces 
 satisfying,
with a given law of evolution  $U(\tau)$, the properties $(1.2),\,(1.3),
\, (1.4)$ has a foliation into a one-parameter (which we shall denote as $s$)
family of isomorphic Hilbert spaces, which are called auxiliary
Hilbert spaces, ${\cal H}_s$ for which
$$ \overline{\cal H} = {\int_\oplus} {\cal H}_s. \eqno(1.8)$$
Representing these spaces in terms of square-integrable functions, we define
the norm in the direct integral space (we use Lebesgue measure)
as 
$$ \Vert f \Vert^2 = \int_{-\infty}^\infty ds \Vert f_s\Vert^2_H, \eqno(1.9)$$
where $f \in {\overline H} $ represents a vector in ${\overline{\cal H}}$
 in terms of a function in the
$L^2$ function space ${\overline H}=L^2(-\infty, \infty, H)$;  $f_s$ is 
an element of $H$, the $L^2$ 
function space  (which we shall call the {\it auxiliary space})
representing ${\cal H}_s$  for any $s$ [we shall not add in what follows a
subscript to the norm or scalar product symbols for scalar products of elements
of the auxiliary Hilbert space associated to a point $s$ on the foliation
axis].
\par The Sinai theorem furthermore asserts that there are representations
for which the action of the full evolution group $U(\tau)$ on 
$L^2(-\infty, \infty;H)$ is translation by $\tau$ units. Given $D_\pm$ (the
subspaces of $L^2$ functions representing ${\cal D}_\pm$), there is such a
representation,
called the {\it incoming representation}$^1$, for which the set of all
functions in $D_-$ have support in $(-\infty, 0)$ and constitute the subspace
 $L^2(-\infty,0;H)$ of $L^2(-\infty, \infty;H)$; there is another
 representation,
called the {\it outgoing representation}, for which functions in $D_+$
have support in $(0,\infty)$ and 
constitute the subspace $L^2(0,\infty;H)$ of $L^2(-\infty, \infty;H)$.
The fact that $Z(\tau)$ in Eq. (1.7) is a semigroup is a consequence of
the definition of the subspaces $D_\pm$ in terms of support properties
on intervals along the foliation axis in the {\it outgoing} and {\it incoming}
translation representations respectively. The non self-adjoint character of
the generator of the semigroup $Z(\tau)$ is a consequence of this structure.
\par Lax and Phillips$^1$ show that there are unitary operators $W_\pm$, 
called wave operators, which map elements in ${\overline{\cal H}}$, 
respectively, to these representations.  They define an $S$-matrix, 
$$ S= W_+W_-^{-1}  \eqno(1.10)$$
which connects the incoming to the outgoing representations; it is
unitary, commutes with 
translations, and maps $L^2(-\infty,0;H)$ into itself. Since $S$
commutes with translations, it is diagonal in Fourier (spectral)
representation.  As pointed out by
Lax and Phillips$^1$, according to a special case of a theorem of
Four\`es and Segal$^{17}$, an operator with these properties can be
represented
as a multiplicative operator-valued function ${\cal S}(\sigma)$ which
maps $H$ into $H$, and satisfies the following conditions:
$$\eqalign{ (a)\   &{\cal S}(\sigma)\  is\  the\  boundary\  
value\  of\  an\ \cr
&operator{\rm -}valued\  function\ {\cal S}(z)\ analytic\ 
for\ {\rm Im}z >0. \cr
(b)\   &\Vert {\cal S}(z) \Vert  \leq 1\ for\ all\ z\ with\  {\rm Im}z >0. \cr
(c)\   &{\cal S}(\sigma)\ is\ unitary\ for\ almost\ all\ real\ \sigma.\cr}$$ 
 An operator with
these properties is known as an inner function$^{18}$; such operators
arise in the study of shift invariant subspaces, the
essential mathematical content of the Lax-Phillips theory. 
 The singularities of
 this $S$-matrix, in what we shall define as the 
{\it spectral representation} (defined in terms of the Fourier
transform on the foliation variable $s$), correspond to the spectrum
of the generator of the semigroup characterizing the evolution 
of the unstable system.    
\par  In the framework of quantum theory, one may identify the Hilbert
space ${\cal H}$ with a space of physical states, and the variable
$\tau$ with the laboratory time (the semigroup
evolution is observed in the laboratory according to this time).
  The representation
of this space in terms of the foliated $L^2$ space ${\overline H}$
provides a natural probabilistic interpretation for the auxiliary spaces 
associated with each value of the foliation variable $s$, i.e., the 
quantity  $\Vert f_s \Vert^2 $ corresponds to the probability density
for the system to be found in the neighborhood of $s$. For example,
consider an operator $A$ defined on ${\overline H}$ which acts pointwise, i.e.,
contains no shift along the foliation. Such an operator can be
represented as a direct integral
$$ A = \int_\oplus A_s.  \eqno(1.11)$$
It produces a map
of the auxiliary space $H$ into $H$ for each value of $s$, and thus,
if it is self-adjoint, $A_s$  may act as an observable in a quantum theory
associated to the point $s$.$^{4}$ The expectation value of $A_s$ in a state
in this Hilbert space defined by the vector $\psi_s$, the component of
$\psi \in {\overline H}$ in the auxiliary space at $s$, is 
$$ \langle A_s \rangle_s = {(\psi_s, A_s \psi_s) \over \Vert \psi_s
\Vert^2}.
 \eqno(1.12) $$
Taking into account the {\it a priori} probability density $\Vert
\psi_s\Vert^2$ that the system is found at this point on the foliation
axis, we see that the expectation value of $A$ in ${\overline H}$ is
$$ \langle A \rangle =  \int ds \langle A_s \rangle_s \Vert \psi_s\Vert^2
= \int ds (\psi_s, A_s \psi_s ), \eqno(1.13)$$    
the direct integral representation of $(\psi, A \psi)$.
\par As we have remarked above, in the translation representations for
 $U(\tau)$ the foliation variable $s$ is shifted (this shift, for
 sufficiently large $\vert \tau \vert$,  induces the transition of
 the state into the subspaces ${\cal D}_\pm$). It follows that $s$
 may be identified as an intrinsic time associated with the evolution
 of the state; since it is a variable of the measure space of the
 Hilbert space ${\overline{\cal H}}$, this quantity itself has the
 meaning of a quantum variable.   
\par We are presented here with the notion of a virtual history. To
understand this idea, suppose that at a given time $\tau_0$, the
function which represents the state has some distribution $\Vert
\psi_s^{\tau_0}\Vert^2$. This distribution provides an {\it a priori}
probability that the system would be found at time $s$ (greater or less than
$\tau_0$), if the experiment were performed at time $s$ corresponding 
 to $\tau = s$ on the laboratory clock.  The state of the system
therefore contains information on the structure of the {\it history} of
 the system as it is inferred at $\tau_0$.   
\par We shall assume the existence of a unitary evolution on the
Hilbert space $\overline{{\cal H}}$, and that for
$$ U(\tau) = e^{-iK\tau}, \eqno(1.14)$$ 
the generator $K$ can be decomposed as 
$$ K= K_0 + V \eqno(1.15)$$
 in terms of an unperturbed
operator $K_0$ with spectrum $(-\infty, \infty)$ and a perturbation
$V$, under which this spectrum is stable.  
We shall, furthermore, assume that wave operators exist, defined on
some dense set, as 
$$ \Omega_\pm = \lim_{\tau \rightarrow \pm \infty} e^{iK\tau}e^{-iK_0
\tau}. \eqno(1.16)$$
In the soluble model that we shall treat as an example in this paper,
 the existence of the wave operators is assured.
\par  With the help of the
 wave operators, we can define translation representations for $U(\tau)$. 
 The translation representation for $K_0$ is defined by the property
$$ {{}_0\langle} s, \alpha \vert e^{-iK_0\tau} f)= {{}_0\langle} s-\tau, \alpha
\vert f), \eqno(1.17)$$
where $\alpha$ corresponds to a label for the basis of the auxiliary space.  
Noting that
$$  K \Omega_\pm = \Omega_\pm K_0 \eqno(1.18) $$
we see that
$$ {{}_{out \atop in}\langle} s, \alpha \vert e^{-iK\tau} f) =
 {{}_{out \atop in} \langle}
s-\tau, \alpha \vert f),  \eqno(1.19)$$
where
$$ {{}_{out \atop in}\langle}s, \alpha \vert f) = {{}_0 \langle}s,
  \alpha \vert 
\Omega_\pm^\dagger f) \eqno(1.20)$$
\par It will be convenient to work in terms of the Fourier transform of
 the {\it in} and {\it out} 
translation representations; we shall call these  the {\it in} and
 {\it out} {\it spectral}
representations, {\it i.e.},
$$ {{}_{out \atop in}\langle} \sigma, \alpha \vert f) = \int_{-\infty}^\infty
e^{-\sigma s} {{}_{out \atop in}\langle} s, \alpha \vert f). \eqno(1.21)$$
In these representations, (1.20) is
$$ {{}_{out \atop in}\langle}\sigma, \alpha \vert f) = {{}_0 \langle}\sigma,
   \alpha \vert \Omega_\pm^\dagger f)
    \eqno(1.22)$$
and $(1.19)$ becomes
$$ {{}_{out \atop in}\langle} \sigma, \alpha \vert e^{-iK\tau}\vert f) =
  e^{-i\sigma \tau}{{}_{out \atop in}\langle} \sigma, \alpha \vert f).
\eqno(1.23)$$
Eq. $(1.17)$ becomes, under Fourier transform
$$ {{}_0\langle} \sigma, \alpha \vert e^{-iK_0\tau} f)= e^{-i\sigma \tau}
{{}_0\langle} \sigma, \alpha \vert f). \eqno(1.24)$$
For $f$ in the domain of $K_0$, $(1.23)$ implies that
$$ {{}_0\langle} \sigma, \alpha \vert K_0 f)= \sigma
{{}_0\langle} \sigma, \alpha \vert f). \eqno(1.25)$$
\par With the solution of $(1.25)$, and the wave operators,
the {\it in} and {\it out} spectral 
representations of a vector $f$ can be constructed from $(1.24)$.
\par We are now in a position to construct the subspaces ${\cal
D}_\pm$, which are not given, {\it a priori}, in the Lax-Phillips quantum
theory.  Identifying ${{}_{out} \langle}s, \alpha \vert f)$ with the  {\it
outgoing} translation representation, we shall define $ D_+$ as the
set of functions with  support in $(0,\infty)$ in this
representation. Similarly, identifying ${{}_{in} \langle}s, \alpha \vert
f)$ with the  {\it
incoming} translation representation, we shall define $ D_-$ as the
set of functions with  support in $(-\infty,0)$ in this
representation. The corresponding elements of ${\overline{\cal H}}$
constitute the subspaces ${\cal D}_\pm$.   By construction,
${\cal  D}_\pm$ have the required invariance properties under the
action of $U(\tau)$.
\par  The {\it outgoing spectral representation}
of a vector $g \in {\cal H}$ is
$$\eqalign{  {{}_{out}\langle} \sigma \alpha \vert g ) =
{{}_0\langle}\sigma\alpha \vert \Omega_+^{-1} g)&= 
\int\,d\sigma' \sum_{\alpha'} {{}_0 \langle} \sigma \alpha \vert {\bf
S}\vert \sigma' \alpha' \rangle_0 {{}_0 \langle}\sigma' \alpha' \vert
\Omega_-^{-1} g)\cr & = \int\,d\sigma' \sum_{\alpha'} {{}_0 \langle} 
\sigma \alpha \vert {\bf
S}\vert \sigma' \alpha' \rangle_0 {{}_{in} \langle}\sigma' \alpha' 
\vert g),\cr} \eqno (1.26)$$
where we call 
$${\bf S} = \Omega_+^{-1} \Omega_-. \eqno(1.27)$$
the quantum Lax-Phillips $S$-operator. We see that the kernel ${{}_0 \langle} 
\sigma \alpha \vert {\bf
S}\vert \sigma' \alpha' \rangle_0$ maps the incoming to outgoing
spectral representations.  Since $\bf S$ commutes with $K_0$, it
follows that 
$$ {{}_0 \langle} 
\sigma \alpha \vert {\bf
S}\vert \sigma' \alpha' \rangle_0 = \delta(\sigma - \sigma') S^{\alpha
\alpha'}. (\sigma) \eqno(1.28)$$
It follows from $(1.16)$ and $(1.22)$, in the standard way$^{19}$, that
$$ {{}_0\langle} \sigma \alpha \vert {\bf S} \vert \sigma' \alpha' 
{\rangle_0} =  \lim_{\epsilon \rightarrow 0}
\delta(\sigma-\sigma')\{\delta^{\alpha \alpha'} - 
2\pi i {{}_0\langle} \sigma \alpha \vert {\bf T}(\sigma + i\epsilon)
 \vert \sigma' \alpha' {\rangle_0} \}, \eqno(1.29)$$
where
$$ {\bf T}(z) = V + VG(z)V = V + VG_0(z) {\bf T}(z). \eqno(1.30)$$
We remark that, by this construction,
$S^{\alpha \alpha'}(\sigma)$ is {\it analytic in the upper half plane} in 
$\sigma$.
The Lax-Phillips $S$-matrix$^1$ is given by the inverse Fourier transform,
$$ S = \bigl\{{{}_0 \langle} 
s \alpha \vert {\bf
S}\vert s' \alpha' \rangle_0\bigr\};  \eqno(1.31)$$
this operator clearly commutes with translations. 
\par From $(1.29)$ it follows that the property $(a)$ above is true. Property 
$(c)$, unitarity for real $\sigma$, is equivalent to asymptotic
completeness, a property which is stronger than the existence of wave
operators. For the relativistic Lee model, which  we shall treat in
this paper, this condition is satisfied.   In the model that we shall 
study here, we shall see that there is a wide class of potentials $V$ for
which the operator $S(\sigma)$ satisfies the property $(b)$ specified above.
\par In the next section, we review briefly the structure of the 
relativistic Lee model$^{19}$, and construct explicitly the
Lax-Phillips spectral representations and $S$-matrix. Introducing
auxiliary space variables, we then characterize the properties of the
finite rank Lee model potential which assure that the $S$-matrix is an
inner function, {\it i.e.}, that property $(b)$ listed above  is satisfied.

\bigskip
{\bf 2. Relativistic Lee-Friedrichs Model}
\smallskip
\par In this section, we define the relativistic Lee model$^{19}$ in terms of
bosonic quantum fields on spacetime ($x \equiv x^\mu$). These fields  evolve
 with an invariant
evolution parameter$^{20}$ $\tau$ (which we identify here with the evolution
 parameter of the Lax-Phillips theory discussed above);
 at equal $\tau$, they satisfy
the commutation relations (with $\psi^\dagger$ as the canonical
conjugate field to $\psi$; the fields $\psi$, which satisfy first order
 evolution equations
as for  nonrelativistic
 Schr\"odinger fields, are just annihilation operators)
$$ [\psi_\tau(x), \psi_\tau^\dagger(x')] =
\delta^4(x-x'). \eqno(2.1)$$
We remark that Antoniou, {\it et el}$^{21}$, have constructed a relativistic
Lee model of a somewhat different type; their field equation is
second order in the derivative with repect to the variable conjugate
to the mass.
\par In momentum space, for which
$$\psi_\tau(p) = {1 \over (2\pi)^2}\int d^4x e^{-ip_\mu x^\mu}
\psi_\tau(x), \eqno(2.2)$$
this relation becomes
$$ [\psi_\tau(p), \psi_\tau^\dagger(p') ] = \delta^4(p-p').\eqno(2.3)$$
The manifestly covariant spacetime structure of these fields is
admissible when $E,{\bf p}$ are not {\it a priori} constrained by a
 sharp mass-shell relation. In the mass-shell limit (for which the
variations in $m^2$ defined by $E^2 - {\bf p}^2$ are small), multiplying both
sides of $(2.3)$ by $\Delta E = \Delta m^2 / 2E$,
 one obtains the usual commutation
relations for on-shell fields,
 $$ [{\tilde \psi}_\tau ({\bf p}), {\tilde \psi}^\dagger({\bf p})] =
2E\delta^3({\bf p} - {\bf p}' ),  \eqno(2.4)$$
where $\tilde \psi ({\bf p}) = \sqrt{\Delta m^2} \psi(p)$.
The generator of evolution 
$$ K= K_0 + V \eqno(2.5)$$       
for which the Heisenberg picture fields  are
$$ \psi_\tau(p) = e^{iK\tau} \psi_0(p) e^{-iK\tau} \eqno(2.6)$$
is given, in this model, as (we write $p^2= p_\mu p^\mu$, $k^2 = k_\mu
k^\mu$ in the following) 
$$\eqalign{K_0 = \int\, d^4p\bigl\{ {p^2 \over 2M_V} b^\dagger(p) b(p) &+ {p^2
\over 2M_N} a^\dagger_N(p) a_N(p) \bigr\} \cr
& +\int \, d^4 k { k^2 \over 2M_\theta } a_\theta^\dagger(k)
a_\theta(k) \cr}\eqno(2.7)$$  
and
$$ V = \int d^4p \int d^4k (f(k) b^{\dagger}(p) a_N(p-k) a_\theta(k) +
   f^*(k) b(p) a^{\dagger}_N(p-k) a^{\dagger}_\theta(k),
   \eqno(2.8)$$
describing the process $V \leftrightarrow N + \theta $. 
The fields $b(p),\, a_N(p)$ and $a_\theta$ are annihilation operators
for the  $V,\, N$, and $\theta$ particles, respectively and $M_v,M_N$ and
$M_\theta$ are the mass parameters for the fields$^{22}$.
\par  The operators $$\eqalign{ Q_1 &= \int \,d^4p [b^\dagger(p)b(p) +
a_N^\dagger(p)a_N(p)]\cr Q_2 &= \int \, d^4p [a_N^\dagger(p)a_N(p) -
a_\theta^\dagger(p) a_\theta(p) ] \cr} \eqno(2.9)$$
are conserved, enabling us to decompose the Fock space to sectors. We
shall study the problem in the lowest sector $Q_1=1, Q_2 = 0$, for
which there is just one $V$ {\it or} one $N$ and one $\theta$. In this sector
the generator of evolution $K$ can be rewritten in the form
$$K=\int d^4p K^p = \int d^4p (K^p_0+V^p)$$
where
$$K^p_0= {{p^2} \over 2M_V}b^\dagger(p)b(p)+ \int d^4k
  \left( {{(p-k)^2}\over 2M_N}+{{k^2} \over 2M_\theta} \right) a^\dagger_N
  (p-k)a^\dagger_\theta(k)a_\theta(k)a_N(p-k)$$
and
$$ V^p = \int d^4k \left(f(k) b^{\dagger}(p) a_N(p-k) a_\theta(k) +
   f^*(k) b(p) a^{\dagger}_N(p-k) a^{\dagger}_\theta(k)\right) $$
In this form it is clear that both $K$ and $K_0$ have a direct
integral structure. This implies a similar structure for the wave operators
$\Omega_{\pm}$ and the possibility of defining restricted wave operators 
$\Omega^p_{\pm}$ for each value of $p$. From the expression for $K_0^p$
we see that $\vert V(p) \rangle =b^\dagger(p)\vert 0 \rangle$ can be regarded
as a discret eigenstate of $K_0^p$ and, therefore, is anihilated by
$\Omega^p_{\pm}$. This, in turn, implies that
$\Omega_{\pm}\vert V(p) \rangle=0$ for every $p$
(an explicit demonstration of this fact is given in appendix A).
\par In order to construct the Lax-Phillips incoming and outgoing
spectral representations for the model presented here it is necessary,
according to the discussion following equation (1.25), to obtain
appropriate expressions for the wave operators $\Omega_\pm^\dagger$ and
the solutions of equation (1.25).
\par We will consider first the problem of finding the generalized
eigenvectors with spectrum $\{\sigma \}$ on $(-\infty,\infty)$,
 $\vert \sigma,\gamma \rangle_0$ of $K_0$. The complete set of
these states is decomposed into two subsets corresponding to the quantum
numbers for states containing $N$ and $\theta$ particles and those 
containing a $V$ particle. These quantum numbers are denoted $\sigma,\alpha$
($N+\theta$ states) and $\sigma,\beta$ ($V$ states) respectively. We have
for the projections into these two subspaces,
$$ \eqalign{ \vert \sigma,\alpha \rangle_0&= \int d^4p
   \int d^4k \vert N(p),\theta(k) \rangle \langle N(p),\theta(k)
   \vert \sigma ,\alpha \rangle_0 \cr
   \vert \sigma,\beta \rangle_0 &= \int d^4p \vert V(p)
   \rangle \langle V(p) \vert \sigma,\beta \rangle_0 \cr}
   \eqno(2.10) $$
where we have denoted $\vert N(p),\theta(k)\rangle \equiv a^\dagger_N(p)
a^\dagger_\theta(k) \vert 0 \rangle$ and $ \vert V(p) \rangle \equiv
b^\dagger(p) \vert 0 \rangle $. Defining
$$ \eqalign{ O^{\sigma,\alpha}_{p,k} &\equiv
   \langle N(p),\theta(k) \vert \sigma,\alpha \rangle_0 \cr
   O^{\sigma,\beta}_p &\equiv \langle V(p)\vert \sigma,\beta
   \rangle_0 \cr}
   \eqno(2.11)$$
we can write (2.10) as
$$ \eqalign{\vert \sigma,\alpha \rangle_0
   &=\int d^4p\,d^4k\,O^{\sigma,\alpha}_{p,k}\vert N(p),\theta(k) \rangle \cr
   \vert\sigma,\beta\rangle_0&=\int d^4p\, O^{\sigma,\beta}_p
   \vert V(p) \rangle \cr}
   \eqno(2.12)$$
It follows from equations (1.25) and (2.12) that we must have
$$ \eqalign{K_0\vert\sigma,\alpha \rangle_0&= \left(
   \omega_{N(p)}+\omega_{\theta(k)}
   \right)\vert \sigma,\alpha \rangle_0 = \sigma \vert \sigma,\alpha
   \rangle_0 \cr
   K_0\vert \sigma,\beta \rangle_0 &= \omega_{V(p)} \vert \sigma,\beta
   \rangle_0 = \sigma \vert \sigma,\beta \rangle_0 \cr}$$
where $ \omega_{N(p)}=p^2/2M_N$,$\omega_{\theta(k)}=
k^2/2M_\theta $ and $ \omega_{N(p)}=p^2/2M_N$.
The kinematic conditions imposed by these equations are satisfied if we set
$$ \eqalign{ O^{\sigma,\alpha}_{p,k}
   &=\delta( \sigma- \omega_{N(p)}
   -\omega_{\theta(k)}) \tilde O^{\sigma,\alpha}_{p,k} \cr
  O^{\sigma,\beta}_p&=\delta( \sigma- \omega_{V(p)})
  \tilde O^{\sigma,\beta}_p \cr}
  \eqno(2.13) $$

A more detailed analysis of the structure of the matrix elements
(2.11) can only be achieved with further knowledge of the nature of
the variables $\alpha$ (defining the auxiliary Hilbert space).
We will postpone the discussion of this point to later and remark here only
that orthogonality and completeness requires that
$$ \eqalign{\sum_\alpha \int d\sigma \left( O^{\sigma,\alpha}_{p,k} \right)^*
   O^{\sigma,\alpha}_{p',k'}&= \delta^4(p-p')\delta^4(k-k') \cr
   \int d^4p\, d^4k\,  \left( O^{\sigma,\alpha}_{p,k} \right)^*
   O^{\sigma^\prime,\alpha^\prime}_{p,k}&=
   \delta(\sigma-\sigma^\prime)\delta_{\alpha,\alpha^\prime} \cr }
   \eqno(2.14) $$
$$ \eqalign{\sum_\beta \int d\sigma \left( O^{\sigma,\beta}_p \right)^*
   O^{\sigma,\beta}_{p'} &= \delta^4(p-p') \cr
   \int d^4p\, \left( O^{\sigma,\beta}_p \right)^*
   O^{\sigma^\prime,\beta^\prime}_p &=
   \delta(\sigma-\sigma^\prime)\delta_{\beta,\beta^\prime} \cr }
   \eqno(2.14') $$

\par We turn now to the calculation of matrix elements of the wave operators
and start by obtaining appropriate expressions for the following matrix
elements of $\Omega_+$
$$ \langle V(p+k) \vert \Omega_+ \vert N(p),\theta (k) \rangle \qquad
   \langle N(p^\prime),\Theta (k^\prime) \vert \Omega_+ \vert N(p),\theta(k)
   \rangle$$
Equation (1.16) can be rewritten, following the standard manipulation $^{23}$,
in the integral form
$$ \Omega_+= 1+i\lim_{\epsilon \to 0} \int_0^{+\infty} U^{\dagger}(\tau)
   VU_0(\tau)e^{-\epsilon \tau} d\tau
   \eqno(2.15)$$
where $U(\tau)=e^{-iK\tau}$, $U_0(\tau)=e^{-iK_0\tau}$. Using (2.7), we have
$$ \Omega_+ \vert N(p),\theta(k) \rangle = \vert N(p),\theta(k) \rangle
   + i\lim_{\epsilon\to 0} \int_0^{+\infty} d\tau U^{\dagger}(\tau)V
   U_0(\tau)e^{-\epsilon\tau} a^{\dagger}_N(p) a^{\dagger}_{\theta}(k)
   \vert 0 \rangle$$
$$ = \vert N(p),\theta(k) \rangle +i \lim_{\epsilon\to 0} \int_0^{+\infty}
   d\tau U^{\dagger}(\tau)Ve^{-i(\omega_N(p)+\omega_\theta(k)-i\epsilon)\tau}
   a^{\dagger}_N(p) a^{\dagger}_\theta(k) \vert 0 \rangle
   \eqno(2.16) $$
Using (2.8) we find 
$$ V a^{\dagger}_N(p)a^{\dagger}_\theta(k) \vert 0 \rangle =
   f(k) b^{\dagger}(p+k) \vert 0 \rangle
   \eqno(2.17) $$
Inserting (2.17) into (2.16) and changing the integration variable from
$\tau$ to $-\tau$ it follows that
$$ \Omega_+ \vert N(p),\theta(k) \rangle = $$
$$ = \vert N(p),\theta(k) \rangle -i \lim_{\epsilon\to 0}
   \int_0^{-\infty} d\tau e^{i(\omega_N(p)+\omega_\theta(k)-i\epsilon)\tau}
   U(\tau)f(k)b^{\dagger}(p+k)\vert 0 \rangle.
   \eqno(2.18)$$

In order to complete the evaluation of the integral in (2.18) we need to
find the time evolution of a state $\psi_0$ under the action of $U(\tau)$
$$ \psi(\tau)=e^{-iK\tau}\psi_0
   \eqno(2.19)$$
In the sector of the Fock space that we are considering, the state
$\psi(\tau)$ at any time $\tau$ can be expanded as
$$ \psi(\tau)=\int d^4q A(q,\tau)b^{\dag}(q) \vert 0 \rangle + \int d^4
   p^\prime \int d^4k^\prime B(p^\prime,k^\prime,\tau)a^{\dagger}_N(p^\prime)
   a^{\dagger}_\theta(k^\prime) \vert 0 \rangle
   \eqno(2.20).$$
In particular, we see that the initial conditions for the evolution in (2.18)
are
$$ B(p',k',0)=0 \qquad A_{pk}(q,0) = f(k)\delta^4(q-p-k) 
   \eqno(2.21) $$
The equations of evolution for the coefficients $A_{pk}(q,\tau)$ and
$B(p',k',\tau)$ are then obtained from (2.19) and (2.20), {\it i.e.},
$$ \eqalign{&i{{\partial A_{pk}(q,\tau)}\over \partial \tau}=
   {{q^2} \over 2M_V} A_{pk}(q,\tau) + \int d^4k' f(k') B(q-k',k',\tau) \cr
   &i{{\partial B(p',k',\tau)} \over \partial \tau}=
   \left( {{{p'}^2}\over
   2M_V}+{{{k'}^2}\over2M_\theta}\right)B(p,'k',\tau)+f^*(k')
   A_{pk}(p'+k',\tau) \cr}
   \eqno(2.22)$$
These equations can be solved algebraically$^{19}$ by performing Laplace
transforms and defining
$$ \eqalign{\tilde A_{pk}(q,z)&= \int_0^{-\infty}
   e^{iz\tau}A_{pk}(q,\tau) d\tau \qquad {\rm Im}z<0 \cr
   \tilde B(p',k',z) &= \int_0^{-\infty}
   e^{iz\tau}B(p',k',\tau) d\tau \qquad {\rm Im}z<0 \cr}
   \eqno(2.23) $$
equations (2.22) are then transformed into
$$ \eqalign{ &\left(z-{{q^2} \over 2M_V} \right) \tilde A_{pk}(q,z)
   =iA_{pk}(q,0)+\int d^4k' f(k') \tilde B(q-k',k',z) \cr
   &\left(z-{{{p'}^2} \over 2M_N}-{{{k'}^2}\over 2M_\theta} \right)
   \tilde B(p',k',z) = iB(p',k',0) + f^*(k) \tilde A_{pk}(p'+k',z) \cr}
   \eqno(2.24)$$
Using the initial conditions (2.21) one obtains the following expressions for
the Laplace transformed coefficients  
$$ \eqalign{ &\tilde A_{pk}(q,z) = {{iA_{pk}(q,0)}\over h(q,z)} \cr
   &\tilde B(p',k',z)=\left(z-{{{p'}^2}\over 2M_N}-
   {{{k'}^2}\over 2M_\theta}\right)^{-1}f^*(k') {{iA_{pk}(p'+k',0)}\over
   h(p'+k',z)} \cr}
   \eqno(2.25) $$
where
$$ h(q,z) \equiv z-{{q^2}\over 2M_V}-\int d^4k { {\vert f(k)\vert^2} \over
   z- {{(q-k)^2}\over 2M_N} -{{k^2}\over 2M_\theta}}
   \eqno(2.26)$$
The Laplace transform of $\psi(\tau)$ is then
$$ \tilde \psi(z)= i \int d^4q {{A_{pk}(q,0)}\over h(q,z)}
   b^{\dagger}(q) \vert 0\rangle $$
$$ +i\int d^4p^\prime \int d^4k^\prime \left(z-{{{p^\prime}^2}\over 2M_N}-
   {{{k^\prime}^2}\over 2M_\theta}\right)^{-1}{{f^*(k^\prime)A_{pk}(p^\prime+
   k^\prime,0)}\over h(p^\prime+k^\prime,z)}
   a^{\dagger}_N(p^\prime)a^{\dagger}_\theta (k^\prime) \vert 0 \rangle
   \eqno(2.27)$$
From (2.18),(2.21) and (2.27) we get
$$ \Omega_+ \vert N(p),\theta(k) \rangle = \vert N(p),\theta (k) \rangle
   +i\lim_{\epsilon\to 0} \Bigl\lbrack -i{{f(k)}
   \over h(p+k,\omega-i\epsilon)}b^{\dagger}(p+k)\vert 0\rangle $$
$$ -i\int d^4p^\prime \left( \omega-i\epsilon
   -{{{p^\prime}^2}\over 2M_N}-{{{p+k-p^\prime}^2}\over 2M_\theta}\right)^{-1}
   {{f^*(p+k-p^\prime)f(k)}\over h(p+k,\omega-i\epsilon)}
   a^{\dagger}_N(p^\prime)a^{\dagger}_\theta(p+k-p^\prime)
   \vert 0\rangle \Bigr\rbrack
   \eqno(2.28)$$
where $\omega = \omega_N(p)+\omega_\theta(k)$.
We can now evaluate the desired matrix elements of the wave operator
$$ \langle V(\tilde p) \vert \Omega_+ \vert N(p),\theta(k) \rangle
   = \lim_{\epsilon\to 0}\delta^4(\tilde p-p-k)
   f(k)h^{-1}(\tilde p,\omega -i\epsilon)
   \eqno(2.29)$$
and
$$ \langle N(\tilde p),\theta(\tilde k)\vert \Omega_+\vert N(p),\theta(k)
   \rangle = \delta ^4(\tilde p-p)\delta ^4(\tilde k-k) $$
$$ +i\lim_{\epsilon\to 0} \Bigl\lbrack -i\left( \omega-i\epsilon
   -{{{\tilde p}^2}\over 2M_N}-{{{\tilde k}^2}\over 2M_\theta}\right)^{-1}
   {{f^*(\tilde k)f(k)}\over h(p+k,\omega-i\epsilon)} \Bigr\rbrack
   \delta ^4(\tilde p +\tilde k -k-p)
   \eqno(2.30)$$
To complete the transformation to the {\it outgoing} spectral representation
we have to calculate, according to eq. (1.24), the following quantities
$$ \langle V(\tilde p) \vert \Omega_+ \vert \sigma,\beta\rangle_0 \qquad
   \langle N(\tilde p),\theta(\tilde k) \vert \Omega_+
   \vert \sigma,\beta \rangle_0 $$
and
$$ \langle V(\tilde p) \vert \Omega_+ \vert \sigma,\alpha\rangle_0 \qquad
   \langle N(\tilde p),\theta(\tilde k) \vert \Omega_+
   \vert \sigma,\alpha \rangle_0 $$
From the second of eq. (2.12), the discussion following eq. (2.8) and the
results of Appendix A, it is clear that the first two transformation matrix
elements are identically zero (since $\Omega_+ \vert V(p)\rangle=0$). For the
first of the last pair we have
$$ \langle V(\tilde p) \vert \Omega_+ \vert \sigma,\alpha \rangle_0 =\int
   d^4p \int d^4k \langle V(\tilde p)\vert \Omega_+ \vert N(p),\theta(k)
   \rangle \langle N(p),\theta(k) \vert \sigma,\alpha \rangle_0$$
$$ = \int d^4p \int d^4k \delta^4(\tilde p-p-k){{f(k)}\over h(p+k,
   \omega -i\epsilon)} O^{\sigma,\alpha}_{k,p} $$
$$ =h^{-1}(\tilde p,\sigma-i\epsilon)\int d^4k f(k)
   O^{\sigma,\alpha}_{k,\tilde p-k}=h^{-1}(\tilde p,\sigma-i\epsilon)
   F^\alpha(\tilde p,\sigma)
   \eqno(2.31) $$
where we have used Eq. (2.13) and the definition 
$$ F^\alpha(\tilde p,\sigma)\equiv \int d^4k f(k)
   O^{\sigma,\alpha}_{k,\tilde p-k}
   \eqno(2.32) $$
Eq. (2.32) can be inverted to find $f(k)$ when given
$F^\alpha(\tilde p,\sigma)$:

$$ \int d^4p_2 \sum_\alpha \int\ d\sigma
   O^{* \sigma,\alpha}_{p_1,p_2} F^\alpha(\tilde p,\sigma)
   =\int d^4p_2 \sum_\alpha \int d\sigma \int d^4k f(k)
   O^{* \sigma,\alpha}_{p_1,p_2} O^{\sigma ,\alpha}_{k,\tilde p-k} $$
$$ =\int d^4 p_2 \int d^4k f(k) \delta^4(\tilde p-k-p_2)\delta^4(k-p_1)
   =f(p_1) \eqno(2.32') $$
We also obtain, in a way similar to the calculation of Eq. (2.31), the second
non-vanishing matrix element

$$ \langle N(\tilde p)+\theta(\tilde k)\vert \Omega_+\vert \sigma,\alpha
   \rangle_0=$$
$$ =\int d^4p \int d^4k \langle N(\tilde p),\theta(\tilde k) \vert \Omega_+
   \vert N(p),\theta(k)\rangle \langle N(p),\theta(k) \vert \sigma,\alpha
   \rangle_0 =O^{\sigma,\alpha}_{\tilde k,\tilde p} $$
$$ +i\lim_{\epsilon\to 0} \Bigl\lbrack -i\left(\sigma-i\epsilon
   -{{{\tilde p}^2}\over 2M_N}-{{{\tilde k}^2}\over 2M_\theta}\right)^{-1}
   f^*(\tilde k)h^{-1}(\tilde p+\tilde k,\sigma-i\epsilon)
   F^\alpha(\tilde p+\tilde k,\sigma) \Bigr\rbrack
   \eqno(2.33) $$
Next we turn to the calculation of the transformation to the {\it incoming}
spectral representation. For this we have to find matrix elements of the
operator $\Omega_-$. The calculation is similar to the one given above
for $\Omega_+$. The integral form of $\Omega_-$ is
$$ \Omega_- = 1+i\lim_{\epsilon\to 0}\int_0^{-\infty} U^{\dag}(\tau)VU_0(\tau)
   e^{\epsilon \tau} d \tau
   \eqno(2.34)$$
For states containing $N$ and $\theta$ particles we have (As for $\Omega_+$
we have that $\Omega_-\vert V(p) \rangle=0$)
$$ \Omega_- \vert N(p),\theta(k) \rangle = \vert N(p),\theta(k) \rangle
   +i\lim_{\epsilon\to 0} \int_0^{-\infty} d\tau U^{\dag}(\tau)VU_0(\tau)
   e^{\epsilon \tau} a^{\dag}_N(p) a^{\dag}_\theta(k) \vert 0 \rangle $$
$$ =\vert N(p)+\Theta(k)\rangle -i\lim_{\epsilon\to 0} \int_0^{+\infty} d\tau
   U(\tau)V e^{i(\omega_N(p)+\omega_\theta(k)+i\epsilon)\tau}
   a^{\dag}_N(p) a^{\dag}_\theta(k) \vert 0 \rangle
   \eqno(2.35) $$
Defining the transformed evolution coefficients for the linear superposition
in this case to be (note the difference from eq. (2.22) )
$$ \eqalign{&\tilde A(q,z)
   =\int_0^{+\infty} d\tau e^{iz\tau}A(q,\tau) \qquad {\rm Im} z > 0\cr 
   &\tilde B(p',k',z)
   = \int_0^{+\infty} d\tau e^{iz\tau}B(p',k',\tau) \qquad
   {\rm Im} z > 0 \cr }
   \eqno(2.36)$$
we find for the matrix elements of the transformation to the
{\it incoming} spectral representation
$$ \langle V(\tilde p) \vert \Omega_- \vert \sigma,\alpha \rangle_0 
   =h^{-1}(\tilde p,\sigma +i\epsilon) F^\alpha(\tilde p,\sigma)
   \eqno(2.37)$$
and
$$ \langle N(\tilde p),\theta(\tilde k)\vert \Omega_-\vert \sigma,\alpha
   \rangle_0 = O^{\sigma,\alpha}_{\tilde k,\tilde p} $$
$$ +i\lim_{\epsilon\to 0} \Bigl\lbrack -i\left(\sigma+i\epsilon
   -{{{\tilde p}^2}\over 2M_N}-{{{\tilde k}^2}\over 2M_\theta}\right)^{-1}
   f^*(\tilde k)h^{-1}(\tilde p+\tilde k,\sigma+i\epsilon)
   F^\alpha(\tilde p+\tilde k,\sigma) \Bigr\rbrack
   \eqno(2.38)$$
This completes the calculation of the Lax-Phillips wave operators providing
the transformations to the {\it incoming} and {\it outgoing} (spectral)
representations. Given these transformations it is possible in principle to
construct the subspaces ${\cal D}_\pm$ according to the method described in
the introduction (see the discussion following eq. (1.25) ). Since the
orthogonality of the resulting subspaces is guaranteed only if the
Lax-Phillips S-matrix satisfies the conditions (a),(b),(c) given in the
introduction, we will calculate it first and then return to a discussion of
the subspaces ${\cal D}_\pm$.
\par From Eq. (1.23) and (1.22) we have
$$ {{}_0}\langle \sigma',\alpha' \vert {\bf S} \vert \sigma,\beta \rangle_0=
   \langle {{}_0}\sigma',\alpha' \vert \Omega_+^{\dagger} \Omega_-
   \vert \sigma,\alpha \rangle_0
   = \int d^4p\, {{}_0}\langle \sigma^\prime ,\alpha^\prime \vert
   \Omega_+^{\dagger} \vert V(\tilde p) \rangle
   \langle V(\tilde p) \vert \Omega_- \vert \sigma ,\alpha \rangle_0 $$
$$ +\int d^4p \int d^4k\,{{}_0}\langle \sigma',\alpha' \vert
   \Omega_+^{\dagger}
   \vert N(\tilde p),\theta(\tilde k) \rangle \langle N(\tilde p),\theta
   (\tilde k) \vert \Omega_- \vert \sigma,\alpha \rangle_0
   \eqno(2.39) $$
Using the expressions obtained for the wave operators Eqs. $(2.32),\,(2.33),
\,(2.37),\,(2.38)$ and the definition (2.31) we find that
$$ {{}_0}\langle \sigma',\alpha' \vert {\bf S} \vert \sigma,\alpha\rangle_0=
   \int d^4\tilde p {{ {F^{\alpha'}}^*(\tilde p,\sigma^\prime)
   F^{\alpha}(\tilde p,\sigma)}\over h(\tilde p,\sigma^\prime+i
   \epsilon)h(\tilde p,\sigma+i\epsilon)}
   +\delta(\sigma^\prime - \sigma) \delta_{\alpha^\prime ,\alpha}$$
$$ +\Bigl\lbrack \left(\sigma+i\epsilon
   -\sigma^\prime \right)^{-1} \int d^4\tilde p
   {{F^\alpha(\tilde p,\sigma){F^{\alpha'}}^*(\sigma^\prime, \tilde p)}
   \over h^{-1}(\tilde p,\sigma+i\epsilon)}
    \Bigr\rbrack
   +\Bigl\lbrack (\sigma^\prime+i\epsilon-\sigma )^{-1}
   \int d^4\tilde p 
   {{{F^{\alpha'}}^*(\tilde p,\sigma^\prime) F^\alpha(\tilde p,\sigma)}
   \over h^{-1}(\tilde p, \sigma^\prime+i\epsilon) }
   \Big\rbrack $$
$$ +\int d^4\tilde p \int d^4\tilde k 
   \Bigl\lbrack \left(\sigma+i\epsilon
   -{{{\tilde p}^2}\over 2M_N}-{{{\tilde k}^2}\over 2M_\theta}\right)^{-1}
   {{\vert f(\tilde k) \vert^2}\over h(\tilde p+\tilde k,\sigma+i\epsilon)}
   F^\alpha(\tilde p+\tilde k,\sigma) \Bigr\rbrack $$
$$ \times \Bigl\lbrack \left(\sigma^\prime+i\epsilon-{{\tilde p^2}
   \over 2M_N}-{{\tilde k^2}\over 2M_\theta} \right)^{-1}
   {{{F^{\alpha'}}^* (\tilde p+\tilde k,\sigma^\prime) }
   \over h^(\tilde p +\tilde k, \sigma^\prime+i\epsilon)} \Big\rbrack
   \eqno(2.40) $$
The last term in eq. (2.40) can be put into a simpler form
by the following manipulation
$$ \int d^4\tilde p  
   \Bigl\lbrack 
   {{F^\alpha(\tilde p,\sigma){F^{\alpha'}}^* (\tilde p,\sigma^\prime)}
   \over h(\tilde p,\sigma+i\epsilon)h(\tilde p, \sigma^\prime+i\epsilon)}
   \Bigr\rbrack $$
$$ \times \int d^4 \tilde k \Big\lbrack
   \left(\sigma+i\epsilon
   -{{{(\tilde p-\tilde k)}^2}\over 2M_N}-
   {{{\tilde k}^2}\over 2M_\theta}\right)^{-1}
   \left(\sigma^\prime+i\epsilon-{{(\tilde p-\tilde k)^2}
   \over 2M_N}-{{\tilde k^2}\over 2M_\theta} \right)^{-1}
   {\vert f(\tilde k) \vert^2}
   \Big\rbrack $$
$$ =\int d^4\tilde p  
   \Bigl\lbrack 
   {{F^\alpha(\tilde p,\sigma){F^{\alpha'}}^* (\tilde p,\sigma^\prime)}
   \over h(\tilde p,\sigma+i\epsilon)h(\tilde p, \sigma^\prime+i\epsilon)} 
   \times {1 \over \sigma^\prime - \sigma}
   \left( \sigma - \sigma^\prime +h(\tilde p,\sigma^\prime +i\epsilon)
   -h(\tilde p,\sigma+i\epsilon) \right) \Bigr\rbrack $$
$$ =-\int d^4\tilde p  
   { {F^\alpha(\tilde p,\sigma){F^{\alpha'}}^* (\tilde p,\sigma^\prime)}
   \over h(\tilde p,\sigma+i\epsilon)h(\tilde p, \sigma^\prime+i\epsilon)}$$
$$ +{1 \over \sigma^\prime - \sigma}\int d^4\tilde p  
   { {F^\alpha(\tilde p,\sigma){F^{\alpha'}}^* (\tilde p,\sigma^\prime)}
   \over h(\tilde p,\sigma+i\epsilon)}
   -{1 \over \sigma^\prime - \sigma}\int d^4\tilde p  
   { {F^\alpha(\tilde p,\sigma){F^{\alpha'}}^* (\tilde p,\sigma^\prime)}
   \over h(\tilde p, \sigma^\prime+i\epsilon)}
   \eqno(2.41) $$
where we have performed a partial fraction decomposition at the second step
in (2.41) and used the definition Eq. (2.26) of $h(q,z)$. Combining (2.41)
and (2.40) we find for the Lax-Phillips $S$-matrix (we use a $i\epsilon$
prescription for dealing with the singularity in (2.41) )
$$ {{}_0}\langle \sigma',\beta' \vert S \vert \sigma,\alpha \rangle_0
   = \delta (\sigma'-\sigma) \Bigl\lbrack \delta_{\alpha^\prime,\alpha}
   -2\pi i \int d^4\tilde p
   { {F^\alpha(\tilde p,\sigma){F^{\alpha'}}^* (\tilde p,\sigma)}
   \over h(\tilde p,\sigma+i\epsilon)} \Big\rbrack
   \eqno(2.42) $$
We observe that in eq. (2.42) the quantity ${F^\alpha}^*(\tilde p,\sigma)$
can be considered, for each fixed value of $\tilde p$, as a vector--valued
function defined on the independent variable $\sigma$, taking its values in
an auxiliary Hilbert space defined by the variables $\alpha$. This observation
allows us to write (see Eqs. (2.11) and (2.32))
$$ {F^\alpha}^*(\tilde p,\sigma)=(\vert n \rangle_{\tilde p,\sigma} )^\alpha
  \eqno(2.43) $$
where (for a fixed value of $\tilde p$) $(\vert n \rangle_{\tilde p,\sigma})
^\alpha$ is the $\alpha$ component of the vector valued function $ \vert n
\rangle_{\tilde p,\sigma}$. With this notation we have (we supress the
auxiliary Hilbert space variables $\alpha$)
$$ S(\sigma) = 1-2\pi i \int d^4\tilde p
   { {\vert n \rangle_{\tilde p,\sigma} 
   \langle n \vert_{\tilde p,\sigma}}
   \over h(\tilde p,\sigma+i\epsilon)} \Big\rbrack
   \eqno(2.44) $$
\par The $S$-matrix given in eq. (2.44) can be simplified after further
investigation into the nature of the auxiliary Hilbert space for this model,
that is, after determination of the auxiliary space variables. This task is
performed in the next section and we find, as expected, the foliation
mentioned in the discussion following eq. (2.9) on the center of momentum
variables. In section 4 we return to further analysis of the
Lax-Phillips S-matrix.
\vfill
\eject
\bigskip
{\bf 3. The auxiliary Hilbert space }
\smallskip
\par The characterization of the auxiliary Hilbert space of the Lax-Phillips
representation of the relativistic Lee-Phriedrichs model is complete once we
give exact specification of the variables $\alpha$ in the
transformation matrix element $O^{\sigma,\alpha}_{p,k}$ of eq (2.11a) (the
discussion above of the Lax-Phillips wave operators and $S$-matrix shows that
for the solution of this problem we do not need detailed information about
$O^{\sigma,\beta}_p$). Determination of these variables
proceeds in two steps. We first define new independent variables
$\{p,k\} \to \{ P,p_{rel} \}$ by the following linear combination of $p$ and
$k$
$$ {\rm a.}\qquad  P=p+k \qquad {\rm b.}\qquad
   p_{rel}={{M_\theta p-M_N k}
   \over M_\theta+M_N}
   \eqno(3.1)$$
these correspond to configuration space variables
$$ {\rm a.}\qquad X_{c.m.}={{M_N x_1 +M_\theta x_2}
   \over M_N +M_\theta}\qquad
   {\rm b.}\qquad x_{rel}=x_1-x_2 $$

From eq. (2.13a) we know that
$$ O^{\sigma,\alpha}_{p,k}=\delta \left( \sigma - {{p^2}\over 2 M_N}
   -{{k^2}\over 2M_\theta} \right) \tilde O^{\sigma,\alpha}_{p,k} $$
This implies that
$$ \sigma = {{p^2}\over 2 M_N}+{{k^2}\over 2M_\theta}
   ={{P^2}\over 2M} + {{p_{rel}^2}\over 2 \mu }
   \eqno(3.2) $$
where $ M=M_N+M_\theta$ and $\mu =(M_N M_\theta)/(M_N+M_\theta) $.
We take $\sigma$ and $P$ to be independent variables; then $ p_{rel}^2 $ is
fixed by:
$$ p_{rel}^2=2\mu \left( \sigma - {{P^2}\over 2M} \right) $$
\par To complete the set of independent quantum numbers we then have to find
a complete set of commuting operators that commute with $p^2_{rel}$ and $P$.
Since $p^2_{rel}$ is a Casimir of the Poincar\'e group on the relative
coordinates, the set of commuting operators contains the second
Casimir of the Lorentz group and, possibly, $L^2$ and $L_3$. We denote the
set of quantum numbers corresponding to the latter three operators
collectively by $\gamma$. As a consequence of this analysis we have that
$\{ \sigma,\alpha \} \equiv \{ \sigma,P,\gamma \}$. From (2.13a) and (3.1a)
it follows that
$$ O^{\sigma,\alpha}_{p,k} \equiv O^{\sigma,P,\gamma}_{p,k}
   = \delta \left( \sigma -{{p^2}\over 2M_N}
   -{{k^2}\over 2M_\theta} \right) \delta^4\left( P-p-k \right)
   \hat O^{p_{rel}^2,\gamma}_{p_{rel}}\Big\vert_{p_{rel}^2=2\mu(\sigma-
   P^2/2M) \atop p_{rel}=(M_\theta p-M_N k)/M}
   \eqno(3.3)$$
Inserting this in the definition of $F^\alpha_{\tilde p,\sigma} (\equiv
F^{P,\gamma}(\tilde p,\sigma))$ we get
$$ F^{P,\gamma}(\tilde p,\sigma)
   = \int d^4k f(k) \delta \left( \sigma -{{(\tilde p-k)^2}\over 2M_N}
   -{{k^2}\over 2M_\theta} \right) \delta^4( P-\tilde p )
   \hat O^{p_{rel}^2,\gamma}_{p_{rel}}\Big\vert_{p_{rel}^2=2\mu(\sigma-
   P^2/2M) \atop p_{rel}=M_\theta \tilde p/M-k} $$
$$ =\delta^4( P-\tilde p )     
   \int d^4p_{rel} f(M_\theta P/ M-p_{rel})
   \delta \left( \sigma -P^2/2M-{{p_{rel}}^2/ 2\mu} \right)
   \hat O^{p_{rel}^2,\gamma}_{p_{rel}}
   \eqno(3.4) $$
In eq. (3.4) we see the foliation on the center of momentum discussed in the
Introduction. We therefore define the following $P$ dependent vector valued
function 
$$ \left( \vert n \rangle_{\sigma,P} \right)^\gamma \equiv     
   \int d^4p_{rel} f^*( M_\theta P/ M-p_{rel})
   \delta \left( \sigma -P^2/2M-{{p_{rel}}^2/ 2\mu} \right)
   (\hat O^{p_{rel}^2,\gamma}_{p_{rel}})^*
   \eqno(3.5) $$
so that $ {F^{P,\gamma}}^*(\tilde p,\sigma)=\delta^4(P-\tilde p)
\left( \vert n \rangle_{\sigma,p} \right)^\gamma $. When this form of
$F^\alpha_{\tilde p,\sigma}$ is used in eq. (2.44) for the $S$-matrix,
taking into account Eq. (2.43), we get
$$ \langle \sigma^\prime ,P^\prime ,\gamma^\prime \vert S
   \vert \sigma ,P,\gamma \rangle
   = \delta (\sigma^\prime -\sigma) 
   \delta^4(P^\prime - P )S^{\gamma',\gamma}_P(\sigma)
   \eqno(3.6) $$
where we define the reduced S-matrix, for a specified value $P$ of 
the center of
momentum 4-vector, to be
$$ S_P(\sigma)\equiv 1-2 \pi i 
   {{\vert n \rangle_{\sigma,P} \langle n \vert_{\sigma,P}}
   \over h(P,\sigma+i\epsilon)}
   \eqno(3.7) $$
The results obtained here make it possible to represent the $S$-matrix in a
form that admits further simplification. In order to achieve this form
we consider first the completeness relation (2.14a) for the transformation
matrix elements $O^{\sigma,P,\gamma}_{p_1,p_2}$. Using eq. (3.3) we
obtain
$$ \delta^4 (p_1^\prime-p_1) \delta^4(p_2^\prime-p_2)=
   \delta^4 (P^\prime-P) \delta^4 (p_{rel}^\prime-p_{rel})=
   \sum_\gamma \int d\sigma \int d^4 P O^{* \sigma,P,\gamma}_{p_1^\prime,
   p_2^\prime}O^{\sigma,P,\gamma}_{p_1,p_2}$$
$$ =\sum_\gamma \int d\sigma \int d^4 \tilde P\, \delta
   (\sigma -{{P^\prime}^2\over 2M}-{{p_{rel}^\prime}^2\over 2\mu})
   \delta^4(\tilde P-P^\prime)
   \hat O^{* 2\mu(\sigma-{{\tilde P^2}\over 2M}),\gamma}_{p_{rel}^\prime} $$
$$ \times \delta (\sigma - {P^2\over 2M}-{{p_{rel}}^2\over 2 \mu })
   \delta^4(\tilde P-P)
   \hat O^{2\mu(\sigma-{{\tilde P^2}\over 2M}),\gamma}_{p_{rel}} $$
$$ =\delta^4(P^\prime-P) \delta ({{p_{rel}^\prime}^2\over 2\mu}
   -{{p_{rel}}^2\over 2\mu})
   \sum_\gamma \hat O^{* p_{rel}^2,\gamma}_{p_{rel}^\prime}
   \hat O^{p_{rel}^2,\gamma}_{p_{rel}} $$
We therefore have a new formulation of the completeness relation
$$ \delta^4 (p_{rel}^\prime-p_{rel})=
   \delta ({{p_{rel}^\prime}^2\over 2\mu}
   -{{p_{rel}}^2\over 2\mu})
   \sum_\gamma \hat O^{* p_{rel}^2,\gamma}_{p_{rel}^\prime}
   \hat O^{p_{rel}^2,\gamma}_{p_{rel}}
   \eqno(3.8) $$
We construct the unit operator on the real $\sigma$ axis by taking the vector
valued function $\vert n \rangle_{\sigma,P}$ and, for each value of $\sigma$,
perform an orthogonalization procedure
$$ 1_{H,\sigma}=\left( 1_{H,\sigma}-{{\vert n \rangle_{\sigma,P} \langle n
   \vert_{\sigma,P}} \over {\scriptstyle \sigma,P}\langle n \vert n
   \rangle_{\sigma,P}}\right)
   +{{\vert n \rangle_{\sigma,P} \langle n \vert_{\sigma,P}}
   \over {\scriptstyle \sigma,P}\langle n \vert n \rangle_{\sigma,P}}
   \eqno(3.9) $$
here $1_{H,\sigma}$ denotes the unit operator in the auxiliary Hilbert space
at a point $\sigma$ on the real axis. Multiplying $S_P(\sigma)$, eq. (3.7), by
the unit operator eq. (3.9) we find
$$ S_P(\sigma)= \left( 1_{H,\sigma}-{{\vert n \rangle_{\sigma,P} \langle n
   \vert_{\sigma,P}} \over {\scriptstyle \sigma,P}\langle n \vert n
   \rangle_{\sigma,P}}\right)
   + {{h(P,\sigma+i\epsilon)-2 \pi i {\scriptstyle \sigma ,P}
   \langle n \vert n \rangle_{\sigma,P} }
   \over h(P,\sigma+i\epsilon)}
   {{\vert n \rangle_{\sigma,P} \langle n \vert_{\sigma,P}}
   \over {\scriptstyle \sigma,P}\langle n \vert n \rangle_{\sigma,P}}
   \eqno(3.10) $$
In order to simplify this expression we shall evaluate ${{}_{\sigma,P}}
\langle n \vert n \rangle_{\sigma,P}$ with the help of the definition, eq.
(3.5), and completeness relation (3.8):
$$ {\scriptstyle \sigma ,P}\langle n \vert n \rangle_{\sigma,P}=
   \sum_\gamma \int d^4p_{rel} \int d^4p_{rel}^\prime
   f^*({{M_\theta P}\over M}-p_{rel}^\prime) 
   f({{M_\theta P}\over M}-p_{rel})\delta (\sigma-{{P^2}\over 2M}
   -{{ {p_{rel}^\prime}^2}\over 2 \mu} ) $$
$$ \times \delta (\sigma-{{P^2}\over 2M} -{{p_{rel}^2}\over 2 \mu} )
   \hat O^{* {p_{rel}^\prime}^2,\gamma}_{p_{rel}^\prime}
   \hat O^{p_{rel}^2,\gamma}_{p_{rel}}= $$
$$ \int d^4p_{rel} \vert f({{M_\theta P}\over M}-p_{rel})\vert^2
   \delta (\sigma - {{P^2}\over 2M}-{{p_{rel}^2}\over 2 \mu})=
   \int d^4k \vert f(k) \vert^2 \delta (\sigma - {{(P-k)^2}\over 2M_N}
   -{{k^2}\over 2M_\theta} )
   \eqno(3.11) $$
We compere this result with the following expression for the jump across the
the cut on the real axis of the complex $\sigma$ plane of $h(P,\sigma)$.
Using eq. (2.26) we obtain
$$ h(P,\sigma+i\epsilon)-h(P,\sigma-i\epsilon)=
   2 \pi i\int d^4k \vert f(k) \vert^2 \delta(\sigma-{{(P-k)^2}\over 2M_N}
   -{{k^2}\over 2 M_\Theta} )
   \eqno(3.12) $$
 From eq. (3.10),(3.11) and (3.12) we see that the Lax-Phillips S-matrix
becomes
$$ S_P(\sigma) =
   \left( 1_{H,\sigma}-{{\vert n \rangle_{\sigma,P} \langle n
   \vert_{\sigma,P}} \over {\scriptstyle \sigma,P}\langle n \vert n
   \rangle_{\sigma,P}}\right)
   +{{h(P,\sigma-i\epsilon)}
   \over h(P,\sigma+i\epsilon)}
   {{\vert n \rangle_{\sigma,P} \langle n \vert_{\sigma,P}}
   \over  {\scriptstyle \sigma,P}\langle n \vert n \rangle_{\sigma,P} }
   \eqno(3.13) $$
It is possible to simplify eq. (3.13) even further by analyzing the behavior
of the projection operator $P_{n,P}(\sigma)$ defined by
$$ P_{n,P}(\sigma)\equiv {{\vert n \rangle_{\sigma,P}
   \langle n \vert_{\sigma,P}}
   \over  {\scriptstyle \sigma,P}\langle n \vert n \rangle_{\sigma,P} }
   \eqno(3.14) $$
This is done in the next section.
\bigskip
{\bf 4. Characterization of the operator $P_{n,P}(\sigma)$}
\smallskip
\par The operator valued function $P_{n,P}(\sigma)$ defined in eq. (3.14)
is a projection operator for each value of $\sigma$

$$P_{n,P}(\sigma) P_{n,P} (\sigma) = P_{n,P}(\sigma)
  \eqno(4.1)$$
It is, therefore, a bounded positive opertor on the real $\sigma$ axis
. In order to characterize
$P_{n,P}(\sigma)$ we need several definitions and results from operator
theory on operator valued functions. We give these in Appendix B, where we
prove that $P_{n,P}(\sigma)$ is an {\it outer function}$^{18}$ and $\vert n
\rangle_{\sigma,P}$ must have the form 
$$\vert n \rangle_{\sigma,P} = g_P(\sigma)\vert n \rangle_P
  \eqno(4.2)$$
where $\vert n \rangle_P$ is a normalized basis vector ${{}_P}\langle n \vert
n \rangle_P=1$ 
\par From eq. (4.2) and (3.11) we get

$$ {{}_{\sigma ,P}}\langle n \vert n \rangle_{\sigma,P}=
   \int d^4k \vert f(k) \vert^2 \delta (\sigma - {{(P-k)^2}\over 2M_N}
   -{{k^2}\over 2 M_\Theta} )=\vert g_P(\sigma) \vert^2
   \eqno(4.3) $$
Furthermore, the operator valued function $P_{n,P}(\sigma)$ becomes
$$ P_{n,P}(\sigma)=
   {{\vert n \rangle_{\sigma,P} \langle n \vert_{\sigma,P}}
   \over {\scriptstyle \sigma,P}\langle n \vert n \rangle_{\sigma,P}}
   =\vert n \rangle_P{{}_P}\langle n \vert
   \eqno(4.4) $$
Here the right hand side is a (constant) vector valued function of $\sigma$.
This result implies a further simplification of the Lax-Phillips
$S$-matrix
$$ S_P(\sigma) =
   1_H-\vert n \rangle_P{{}_P}\langle n \vert
   + {{h(P,\sigma-i\epsilon)}
   \over h(P,\sigma+i\epsilon)}
   \vert n \rangle_P{{}_P}\langle n \vert
   \eqno(4.5) $$
\par We now complete the characterization of the Lax-Phillips S-matrix
$S_P(\sigma)$. The projection valued function $P_{n,P}(\sigma)=\vert n
\rangle_P {{}_P}\langle n \vert $ of equation (4.4) is to be thought of as a
realization, as an operator valued function, of an operator $P_{n,P}$ which
projects all functions in $L^2(-\infty,+\infty;H)$ on the subspace
containing all functions taking their values in the one dimensional subspace
of the auxiliary Hilbert space spanned by the single basis vector
$\vert n \rangle_P$. If we denote the subspace of $H$ spaned by $\vert
n \rangle_P$ by $H_1$ then we have
$$ P_{n,P} \colon L^2(-\infty,+\infty;H)\to L^2(-\infty,+\infty;H_1)
   \eqno(4.6)$$
We denote by $P_{1-n,P}$ the operator projecting on the subspace
of functions with a range in $H \ominus H_1$. We have
$$P_{1-n,P}\colon L^2(-\infty,+\infty;H)
  \to L^2(-\infty,+\infty;H \ominus H_1)
  \eqno(4.7)$$
This operator is realized in a simple way on $L^2(-\infty,+\infty;H)$ as the
operator valued function
$$T(P_{1-n,P})=\delta(\sigma-\sigma')(1_H-\vert n \rangle_P{{}_P}\langle n
  \vert)
  \eqno(4.8)$$
(if $A$ is an operator on the Hardy class $H_H^2(\Pi)$ then $T(A$) is a
realization of the operator $A$ is its realization in terms of an operator
valued function). It is obvious from eq. (4.4) and (4.8) that

$$L^2(-\infty,+\infty;H)=P_{n,P}L^2(-\infty,+\infty;H)\oplus P_{1-n,P}
  L^2(-\infty,+\infty;H)
  \eqno(4.9)$$
In particular, if $U(\tau)$ is the operator for right shift by $\tau$ units
(a left shift for $\tau<0$) then any left shift invariant subspace
$I^-_H \subset H^2_H(\Pi)$ ($H^2_H(\Pi)$ is a Hardy class of the upper half
complex $\sigma$ plane, see Appendix B for notation; we are working in
the spectral representations of the Lax-Phillips theory) can be written as
$$I^-_H=P_{n,P}I^-_H\oplus P_{1-n,P}I^-_H
  \eqno(4.10)$$
Now, in the spectral representation the shift operator is represented by
multiplication by $e^{-i\sigma \tau}$ and we have
$$ [U(\tau),P_{n,P}]=[U(\tau),P_{1-n,P}]=0
   \eqno(4.11)$$
Furthermore, $I^-_H$ is a left shift invariant subspace and therefore
$$ U(\tau)I^-_H\subset I^-_H\qquad \tau<0
   \eqno(4.12)$$
Denote $I^-_{H_1}\equiv P_{n,P}I^-_H$, then, taking into account eq. (4.12)
and (4.11) we find the following relation
$$U(\tau)I^-_{H_1}=U(\tau)P_{n,P}I^-_H=P_{n,P}U(\tau)I^-_H\subset P_{n,P}
  I^-_H=I^-_{H_1} \qquad \tau<0$$
or
$$ U(\tau)I^-_{H_1}\subset I^-_{H_1} \qquad \tau<0
   \eqno(4.13)$$
We see that if $I^-_H$ is an invariant subspace for left shifts then
$I^-_{H_1}=P_{n,P}I^-_H$ is a one dimensional invariant subspace under left
shifts.
\par In the Lax-Phillips theory the Lax-Phillips $S$-matrix generates
a left shift invariant subspace from the Hardy class $H_H^2(\Pi)$
(this corresponds to the stability property of ${\cal D_-}$). In this case
we can write
$$ I^-_H=S^{LP} H^2_H(\Pi)
   \eqno(4.14)$$
where $S^{L.P}$ is the Lax-Phillips S-matrix. From eq. (4.4) and (4.5)
we see that in the case of the relativistic Lee-model we have ($S_P(\sigma)$
is the realization of $S^{L.P}$ in terms of an operator valued function)

$$ [S^{LP},P_{n,P}]=0
   \eqno(4.15)$$
From eq. (4.14), (4.15) and the definition of $I^-_{H_1}$ we see that in this
case
$$ I^-_{H_1}=P_{n,P}I^-_H=P_{n,P}S^{L.P}H^2_H(\Pi)=S^{LP}P_{n,P}
   H^2_H(\Pi)=S^{LP}H^2_{H_1}(\Pi)$$
where $H^2_{H_1}(\Pi)\equiv P_{n,P}H^2_H(\Pi)$. This can be rewritten as
$$ I^-_{H_1}=P_{n,P}S^{LP}P_{n,P}H^2_{H_1}(\Pi)
  \eqno(4.16)$$
From this we see that $P_{n,P}S^{LP}P_{n,P}$ generates a one dimensional
left shift invariant subspace from $H_H^2(\Pi)$. From eq. (4.4) and (4.5) we
get

$$ T(P_{n,P}S^{LP}P_{n,P})=\delta(\sigma-\sigma')
   {{h(P,\sigma-i\epsilon)}
   \over h(P,\sigma+i\epsilon)}
   \vert n \rangle_P{{}_P}\langle n \vert
   \eqno(4.17)$$
This immediately implies that $h(P,\sigma-i\epsilon)/h(P,\sigma+i\epsilon)$ is
a {\it scalar inner function}$^{24}$. This result comprises a complete
characterization of $S_P(\sigma)$ in eq. (4.5).
\par A scalar inner function $f$ can always be written as a product $f=RE$,
where $R$ is a rational inner function containing all of the zeros (and poles)
of $f$ and $E$ is called the singular part of $f$ and is an inner function
with no zeros$^{23}$. If we assume that in the case of the relativistic Lee-model
we have only a single pole in the lower half of the complex $\sigma$ plane,
corresponding to a single resonance of the model, then we have in general

$$ {{h(P,\sigma-i\epsilon)} \over h(P,\sigma+i\epsilon)}={{\sigma-
  {\overline\mu}_P}\over \sigma-\mu_P}e^{if_P(\sigma)}
   \eqno(4.18)$$
where $R=(\sigma-\tilde \mu_P)/(\sigma-\mu_P)$ is the rational part of the
inner function and $E=e^{if_P(\sigma)}$ is the singular part.
\par Using eq. (4.18) in the expression for the Lax-Phillips $S$-matrix
eq. (4.5) we note that it can be written in terms of the following product

$$S_P(\sigma)=S'_P(\sigma)M_P(\sigma)
  \eqno(4.19)$$
where
$$ \eqalign{S'_P(\sigma)&= 1_H-\vert n \rangle_P{{}_P}\langle n \vert+
   {{\sigma-{\overline\mu}_P}\over \sigma-\mu_P}\vert n \rangle_P{{}_P}
   \langle n \vert \cr
   M_P(\sigma)&=1_H-\vert n \rangle_P{{}_P}\langle n \vert+e^{if_P(\sigma)} 
   \vert n \rangle_P{{}_P}\langle n \vert \cr}
   \eqno(4.20)$$
The factor $S'_P(\sigma)$ is a rational inner function containing the pole
and zero of $S_P(\sigma)$ and $M_P(\sigma)$ is an inner function with
no zeros. If $M_P(\sigma)$ is of exponential growth, that is, if it true that
$$ \vert M(\sigma) \vert \leq e^{k\vert {\rm Im}\sigma \vert}$$
for all complex $\sigma$ and for some positive $k$, then $M_P(\sigma)$ is
called a {\it trivial inner factor}$^1$.
\par In the framework of the Lax-Phillips theory it is possible to consider
equivalent {\it incoming} and {\it outgoing} representations. These are
defined in the following way$^1$
\smallskip
{\bf Definition:} Two incoming (outgoing) subspaces $D$ and $D'$ are called
{\it equivalent} with respect to the evolution $U(\tau)$ if there exists a
real number $T$ such that
$$U(T)D\subset D' \qquad U(T)D'\subset D$$
\smallskip
Lax and Phillips proved that if $D_-$ and $D'_-$ are equivalent {\it incoming}
subspaces then the spectral representations obtained with respect to $D_-$
and $D'_-$ are connected by a trivial inner factor. If $a$ represents
a vector f in {\cal H} in the spectral representation with respect to $D_-$
and $a'$ represents the same $f$ in the spectral representation with respect
to $D'_-$, then there is a trivial inner factor $M_-(\sigma)$ such that
$$a'(\sigma)=M_-(\sigma)a(\sigma)
  \eqno(4.21)$$
A similar result holds for two equivalent {\it outgoing} subspaces. In this
case the trivial inner factor is denoted by $M_+(\sigma)$. Furthermore, if
$D_- \, , D'_-$ and $D_+ \, ,D'_+$ are pairs of equivalent {\t incoming} and
{\it outgoing} subspaces satisfing, respectively, the assumptions of the
Lax-Phillips theory then the associated scattering matrices are related by
$$ S'=M_+SM_-^{-1},$$
where $M_+$ and $M_-$ are the trivial inner factors which connect the
spectral representations with respect to the two equivalent {\it outgoing}
and {\it incoming} subspaces respectively. Lax and Phillips prove that the
properties of the two scattering matrices are essentially unchanged, i.e.
$S'(\sigma)$ and $S(\sigma)$ are singular at the same points in the complex
plane and the generators of the semigroup in the two cases have the same
spectrum. We see that in the case of equivalent problems, in the sense of the
definition above, we can reduce the problem of finding the resonant state of
$S_P(\sigma)$ to that of finding the resonant state of the equivalent problem
for which the  $S$-matrix is the rational factor $S'_P(\sigma)$ in eq. (4.20)
and then transforming with the trivial inner factor $M_P(\sigma)$ using eq.
(4.19) and (4.21) (or the analogous statement to eq. (4.21) for two
equivalent {\it outgoing} subspaces). In the next section we find the resonant
state for the case of the rational S-matrix $S'_P(\sigma)$ of eq. (4.20). The
particle, {\it i.e.},$N\, ,\theta$ or $V$, content of the resonant state in
this case is calculated explicitly.
\bigskip
{\bf 5. The resonant state for the rational case}
\smallskip
\par In this section we shall identify the resonant state of the relativistic
Lee-model in the Lax-Phillips {\it outgoing} translation representation for
the case of a rational $S$-matrix of the form

$$ S_P(\sigma)= 1_H-\vert n \rangle_P{{}_P}\langle n \vert+
   {{\sigma-\overline \mu_P}\over \sigma-\mu_P}\vert n \rangle_P{{}_P}
   \langle n \vert 
   \eqno(5.1)$$
(this can also be achieved, in the same way, in the {\it incoming}
translation representation). Having found the resonant state, we will use the
transformation given in eq. (2.31) and (2.33) (or (2.37),(2.38) for the
{\it incoming} representation),
to see the particle content of the resonance.
\par In order to identify the resonant state we construct the projection
operator $P_-$ of the Lax-Phillips theory$^1$. The Hilbert space of the
Lax-Phillips theory is decomposed as
$${\cal H} = {\cal D_-} \oplus {\cal K} \oplus {\cal D_+}$$
The operator $P_-$ is the projection into the subspace ${\cal K} \oplus
{\cal D_+}$. In the {\it outgoing} translation representation $D_+$ is
given by $L^2(0,\infty;H)$, i.e. it is defined in a simple way by its
support properties. If we obtain an expression for $P_-$ in this
representation and identify the projection onto $D_+$, then the remaining
part will necessarily be the projection into $K$, the subspace of the
resonance (a similar procedure for the identification the resonance involves
the representation of $P_+$ in the {\it incoming} translation representation
and noting that in this representation $D_-$ is given in terms of support
properties).
\par We use the fact that the subspace $D_-$ is given in the {\it incoming}
translation representation in terms of its support properties. This allows us
to write

$$ P_- = \sum_{\alpha''}\int d\eta\, \vert \eta,\alpha'' \rangle_{in}
   \theta(\eta) {{}_{in}}\langle \eta,\alpha'' \vert
   = \sum_{\alpha''} \int d\eta\, \Omega_- \vert \eta,\alpha''
   \rangle_0 \theta(\eta)
   {{}_0}\langle \eta,\alpha'' \vert \Omega_-^\dagger
   \eqno(5.2)$$
To represent $P_-$ in the {\it outgoing} translation representation we apply
eq. (1.20) and obtain
$$ {{}_{out}}\langle s,\alpha \vert P_-
   \vert s^\prime,\alpha^\prime \rangle_{out} =
   \sum_{\alpha''} \int d\eta\,
   {{}_{out}}\langle s,\alpha \vert
   \Omega_- \vert \eta,\alpha'' \rangle_0 \theta(\eta)
   {{}_0}\langle \eta,\alpha'' \vert \Omega_-^\dagger
   \vert s^\prime,\alpha' \rangle_{out} $$
$$ = \sum_{\alpha''} \int d\eta\,
   {{}_0}\langle s,\alpha \vert \Omega_+^\dagger
   \Omega_- \vert \eta,\alpha'' \rangle_0 \theta(\eta)
   {{}_0}\langle \eta,\alpha'' \vert \Omega_-^\dagger \Omega_+
   \vert s^\prime,\alpha' \rangle_0 $$
$$ = \sum_{\alpha''} \int d\eta\,
   {{}_0}\langle s,\alpha \vert S
   \vert \eta,\alpha'' \rangle_0 \theta(\eta)
   {{}_0}\langle \eta,\alpha'' \vert S^\dagger
   \vert s^\prime,\alpha' \rangle_0 $$
In this expression should to use the spectral representation of the
scattering operator $S$ and its adjoint $S^\dagger$. Performing the proper
Fourier transforms and taking into account eq. (3.6) we get

$$ {\scriptstyle out}\langle s,P,\gamma \vert P_-
   \vert s^\prime,P',\gamma^\prime \rangle_{out}=$$
$$ \delta^4(P-P')\left[\sum_{\gamma''} \int d\sigma \int d\sigma^\prime
   \int d\eta e^{i\sigma s}S_P^{\gamma,\gamma''}(\sigma)e^{-i\eta \sigma}
   \theta(\eta) e^{i\eta \sigma^\prime}
   {S_P^\dagger(\sigma^\prime)}^
   {\gamma'',\gamma'}e^{-i\sigma^\prime s^\prime}\right] $$
$$ =\delta^4(P-P')\left[{{-i}\over 4\pi^2}
   \int d\sigma \int d\sigma^\prime \sum_\alpha e^{i\sigma s}
   {{S_P^{\gamma,\gamma''}(\sigma)
   {S_P^\dagger(\sigma^\prime)}^{\gamma'',\gamma'}}
   \over \sigma - (\sigma^\prime +i\epsilon)}
   e^{-i\sigma^\prime s^\prime}\right]
   \eqno(5.3) $$
The operator valued function $S_P(\sigma)$ is analytic in the upper half
of the complex $\sigma$ plane. The adjoint $S_P^\dagger(\sigma)$ is analytic
in the lower half plane. We assume that $S_P(\sigma)$ is in the form of eq.
(5.1). If the pole of $S_P(\sigma)$ is at the point $\mu_P$,
we have that the pole of $S_P^\dagger(\sigma)$ is at $\overline \mu_P$ and

$$ S_P(\sigma)=1+ {{{\rm Res}S_P(\mu_P)}\over \sigma-\mu_P} \qquad
   S_P^\dagger (\sigma)=1+ {{ {\rm Res}S_P^\dagger (\overline \mu_P) }
   \over \sigma-\overline \mu_P } \qquad {\rm Im}\mu_P < 0
   \eqno(5.4)$$
From eq. (5.4) we see that contour integration is allowed when performing the
integrals in eq. (5.3) for the various signs of $s$ and $s'$. The result is

$$ {{}_{out}}\langle s,P,\gamma \vert P_-
   \vert s^\prime,P',\gamma^\prime \rangle_{out}=\delta^4(P-P')$$
$$ \times \Bigl\{ \theta(s)\delta(s-s^\prime)\delta_{\gamma,\gamma'}
   +{1\over 2\pi} \theta(-s)
   \Bigl[e^{i\mu_P s}{\rm Res}\ S_P(\mu_P) \int d\sigma^\prime
   {{S_P^\dagger(\sigma^\prime)}\over \mu_P-(\sigma^\prime+i\epsilon)}
   e^{-i \sigma^\prime s^\prime} \Bigr]^{\gamma,\gamma'}=$$
$$ \theta(s)\delta(s-s^\prime)\delta_{\gamma,\gamma'}
   -i\theta(-s)\theta(s^\prime)
   \Bigl[e^{i \mu_P s}{\rm Res}\ S_P(\mu_P) S_P^\dagger(\mu_P) 
   e^{-i \mu_P s^\prime}\Bigr]^{\gamma,\gamma'} $$
$$ +i \theta(-s) \theta(-s^\prime)
   \Bigl[ e^{i \mu_P s}{\rm Res}\ S_P(\mu_P) 
   {{ {\rm Res}\ S_P^\dagger(\overline \mu_P)}
   \over \mu_P-\overline \mu_P }
   e^{-i \overline \mu_P s^\prime} \big]^{\gamma,\gamma'}\Bigr\} $$
From eq. (5.1) we have
$$ {\rm Res}\ S(\mu_P)S^\dagger(\mu_P)=0 $$
and as a consequence (in the sequel we will suppress the indices
$\gamma,\,\gamma'$)
$$ {{}_{out}}\langle s,P \vert P_-\vert s^\prime,P'\rangle_{out}=
   \delta^4(P-P')$$
$$ \times \bigl\{ \theta(s)\delta(s-s^\prime)+i \theta(-s) \theta(-s^\prime)
   \Bigl[ e^{i \mu_P s}{\rm Res}\ S_P(\mu_P) 
   {{ {\rm Res}\ S_P^\dagger(\overline \mu_P)}
   \over \mu_P-\overline \mu_P }
   e^{-i \overline \mu_P s^\prime} \big] \big\}
   \eqno(5.5) $$
The first term in eq. (5.5) is the projection into $D_+$ which, in this
representation, is identified with the set of functions $L^2(0,\infty;H)$. The
second term in eq. (5.5) is therefore the projection into the subspace $K$
of the resonant state. With the help of eq. (5.1) we obtain for the second
term of eq. (5.5)
$$ \delta^4(P-P')\bigl[2{\rm Im}\mu_P(\Theta(-s)
   e^{i \mu_P s}\vert n\rangle_P)
   (\theta(-s^\prime)e^{-i \overline \mu_P s^\prime}
   {{}_P}\langle n \vert)\bigr]$$
The resonant state is the eigenstate of this projection operator. If we
denote the resonant state by $\vert R \rangle_P$ then, for a specified value
of $P$, we have
$$ {{}_{out}}\langle s,P,\gamma \vert R \rangle_{P'}=
   \delta^4(P-P')2{\rm Im}\mu(P)
   \theta(-s) e^{i\mu_P s}(\vert n \rangle_P)^\gamma
   \eqno(5.6)$$
(here $2{\rm Im}\mu_P$ is a normalization constant). In the spectral
representation we obtain
$$ {{}_{out}}\langle \sigma,P',\gamma \vert R \rangle_{P}=\delta^4(P-P')
   2i{\rm Im}\mu_P{ {(\vert n \rangle_P)^\gamma} \over \sigma - \mu_P}
   \eqno(5.7)$$
In order to see the particle content of the resonance we calculate the
following transformation
$$\eqalign{\langle N(p),\theta(k) \vert R \rangle_P &= \sum_\alpha \int
  d\sigma \langle N(p),\theta(k) \vert \sigma, \alpha \rangle_{out}
  {{}_{out}}\langle \sigma,\alpha \vert R \rangle_P \cr
  \langle V(p) \vert R \rangle_P &= \sum_\alpha \int d\sigma
  \langle V(p) \vert \sigma,\alpha \rangle_{out} {{}_{out}}\langle
  \sigma,\alpha \vert R \rangle_P \cr}
  \eqno(5.8)$$
For the calculation of the transformations in eq. (5.8) we use Eq. (2.31),
(2.33) and (5.7). Taking into account Eq. (3.3),(3.5),(3.8) and (4.2), the
first term of Eq. (2.33) gives
$$ 2i{\rm Im}\mu_P \sum_\gamma \int d\sigma
   O^{\sigma ,P,\gamma}_{k,p}
   { {(\vert n \rangle_P)^\gamma} \over \sigma - \mu_P}=
   2i{\rm Im}\mu_P \sum_\gamma \int d\sigma\, g^{-1}(\sigma)
   O^{\sigma ,P,\gamma}_{k,p}
   {{(g(\sigma)\vert n \rangle_P)^\gamma} \over \sigma - \mu_P}=$$
$$ 2i{\rm Im}\mu_P{{-g^{-1}(\omega_N(p)+\omega_\theta(k))f^*(k)}
   \over \mu_P-{{p^2}\over 2M_N}-{{k^2}\over 2M_\theta}}
   \delta^4(P-p-k)
   \eqno(5.9)$$
where, as before, $\omega_N(p)=p^2/2M_N$ and $\omega_\theta(k)=k^2/2M_\theta$.
The second term of eq. (2.33) give the contribution

$$ 2i{\rm Im}\mu(P)f^*(k)\delta^4(P-p-k)
   \int d\sigma h^{-1}(P,\sigma-i\epsilon) (\sigma-i\epsilon
   -{{p^2}\over 2M_N}-{{k^2}\over 2M_\theta})^{-1}
   {{g^*_P(\sigma)} \over \sigma - \mu_P}
   \eqno(5.10)$$
Assume that $g^*_P(\sigma)$ is analytic in the lower half of the complex
$\sigma$ plane and it possible to perform countour integretion; then, if
we find that (5.10) reduces to
$$ 4\pi{\rm Im}\mu_P f^*(k)\delta^4(P-p-k)
   {1 \over \mu_P-{{p^2}\over 2M_N}-{{k^2}\over 2M_\theta} }
   {{g^*_P(\mu_P)}\over h(P,\mu_P)}
   \eqno(5.11)$$
In this case we obtain, for the first transformation in eq. (5.8)

$$\langle N(p),\theta(k) \vert R \rangle_P=$$
$$ -2i{\rm Im}\mu_P\delta^4(P-p-k){{f^*(k)}\over
   \mu_P-{{p^2}\over 2M_N}-{{k^2}\over 2M_\theta}}
   \left({{2\pi i g_P(\omega_N(p)+\omega_\theta(k))g^*_P(\mu_P)+h(P,\mu_P)}
   \over h(P,\mu_P)g_P(\omega_N(p)+\omega_\theta(k))}\right)
   \eqno(5.12)$$
For the second transformation in (5.8) we use eq. (2.31) and, assuming
again the same analytic properties of $g^*_P(\sigma)$, we find

$$\langle V(\tilde p) \vert R \rangle_P
  = 2i{\rm Im}\mu_P \delta^4(\tilde p-P){{g^*_P(\mu_P)}\over h(P,\mu_P)}
  \eqno(5.13)$$
\par  Note that if the function $g^*(\sigma)$ is a constant
function then from eq. (4.3) we have $g_P(\omega_N(p)+\omega_\theta(k))
g^*_P(\mu_P)=\vert g_P(\sigma) \vert^2={{}_{\sigma,P}}\langle n \vert n
\rangle_{\sigma,P}=\vert g_P(\mu_P) \vert^2$ in this case the numerator in eq.
(5.12) vanishes and there is no $N,\theta$ component in the resonant
 state.  In the case that the complete $S_P(\sigma)$ is a rational
 function,  $\vert g^*(\sigma)\vert^2$ is a constant (see Appendix C), equal to
 $-(1/\pi) {\rm Im} \mu_P$, as in the approximate  Wigner-Weisskopf
 theory of the Lee-Friedrichs model of resonance$^{10}$.  In this case,
  $g^*(\sigma)$ is itself
 not necessarily a constant, but proportional to a non-trivial phase. 
\bigskip
{\bf 6. Conclusions}
\smallskip
 \par We have studied the application of Lax-Phillips quantum theory
 to a soluble relativistic quantum field theoretical model. In this
 model, we obtain the Lax-Phillips
 $S$-matrix explicitly as an inner function (the Lax-Phillips structure
 is defined pointwise on a foliation over the total energy-momentum of
 the system).  The structure of the Lee model $S$-matrix $(2.42)$ has a
 term with factorized numerator, corresponding to the transition
 matrix element of the interaction, and denominator $h({\tilde p},
 \sigma)$ which contains the zero inducing the resonance pole. The
 numerator factors are identified as a vector field over the complex
 $\sigma$ plane. Foliating the $S$-matrix over the total energy
 momentum $P$, it takes on the form of a projection into the space
 complementary to the discrete subspace of the rank one potential of
 the model (for each point $\sigma,\, P$), plus an scalar inner
 function on the discrete subspace.   
 The vector field on the complex extension of the
 spectral representation (on the singular point, it corresponds to the
 projection into the resonant eigenstate), is proven (Appendix B)
 to be independent
 of the spectral parameter up to a scalar multiplicative function. It
 then follows  that
 the projection is in fact independent of $\sigma$.
 This result leads to the conclusion that the properties of the $S$-matrix are
 essentially derived from the properties of a scalar inner function.   
\par This inner function consists of, in general, a rational factor,
 which contains all of the zeros and poles, and a singular factor
 (constructed with singular measure).  If the singular factor is
 exponentially bounded, it is, in the terminology of Lax and
 Phillips$^1$, a trivial inner function.  The application of this
 inner function does not change the resonance structure, but the
 functional form of the eigenfunctions and scattering states may be
 altered.
\par We then studied the rational case, the simplest possible model
 for a non-trivial Lax-Phillips theory, for which the inner function
 reduces to just the ratio $ (\sigma - {\overline \mu}_P) /(\sigma -
 \mu_P)$.  We therefore see, conversely, that the simplest model for a
 non-trivial Lax-Phillips theory corresponds to a rank one Lee
 model$^{25}$.
\par For the rational case, we find the explicit form $(5.7)$ of the resonance
 state in the outgoing spectral representation, of the same structure as
 given by Lax and Phillips$^1$. We furthermore give a formula for the
 $N,\theta$ and $V$ components of the resonance. In the case that
 $g_P(\sigma)$ is independent of $\sigma$, there is no $N,\theta$
 component.  In Appendix C, we show (for the rational case) that the
 absolute value of
 $g_P(\sigma)$ is independent of $\sigma$, but that there is, in
 general, a phase which admits an $N,\theta$ component in the
 resonance$^{24}$.
\par The case of more than one channel (resonance) for the
 relativistic  Lee model can be treated in a similar way.  In
 particular, for the two channel case, providing a model, for example,
 of the neutral $K$-meson decay, one can show that the
 phenomenological model of Lee, Oehme and Yang$^{12}$ and Wu and
 Yang$^{13}$ can be constructed directly from the more fundamental 
quantum Lax-Phillips theory. This work will be treated in a succeeding
 paper$^{26}$.     
\par The study of the relativistic Lee model that we have given here
 appears to have all of the basic properties of a Lax-Phillips theory
 for a more general quantum field theory. We are presently studying
 the LSZ construction$^{27}$ in this framework. 
\bigskip
\noindent
{\it Acknowledgements\/}
\par We wish to thank S.L. Adler for suggesting the application of the
Lax-Phillips method to quantum field theory, and  E. Eisenberg, E. Gluskin and
C. Piron for useful discussions. One of us (L.H.) also wishes to thank Professor Adler for his hospitality at the Institute for Advanced Study, where this work was completed.
   
\bigskip
{\bf Appendix A.}
\smallskip
\par We show that $\Omega_\pm \vert V(\tilde p) \rangle = 0$ applying the
methods used in section 2. The procedure explicitly performed for
$\Omega_+ \vert V(\tilde p) \rangle = 0$. The result for $\Omega_- \vert
V(\tilde p) \rangle = 0$ in obtained in a similar way.
\par we start with the integral representation of the wave operator (see eq.
(2.15))
$$ \Omega_+= 1+i\lim_{\epsilon \to 0} \int_0^{+\infty} U^{\dagger}(\tau)
   VU_0(\tau)e^{-\epsilon \tau} d\tau
   \eqno(A.1)$$
applying this operator to $\vert V(\tilde p) \rangle$ we get

$$ \Omega_+ \vert V(\tilde p) \rangle = \vert V(\tilde p) \rangle
   + i\lim_{\epsilon\to 0} \int_0^{+\infty} d\tau U^{\dagger}(\tau)V
   U_0(\tau)e^{-\epsilon\tau} b^{\dagger}(\tilde p) \vert 0 \rangle$$
$$ = \vert V(\tilde p) \rangle -i \lim_{\epsilon\to 0} \int_0^{-\infty}
   d\tau U(\tau)Ve^{i(\omega_V(\tilde p)-i\epsilon)}
   b^{\dagger}(\tilde p) \vert 0 \rangle
   \eqno(A.2) $$

\par As in section 2, we want to evaluate the time evolution in the integral
and perform a Laplace transform. The result of the action of the potential
operator, given in eq. (2.8) to $\vert V(p) \rangle$, is

$$ V\vert V(\tilde p) \rangle=Vb^\dagger(\tilde p)\vert 0\rangle=
   \int d^4k f^*(k)a^\dagger_N(\tilde p-k)a^\dagger_\theta(k) \vert 0 \rangle
   \eqno(A.3) $$
A general form of a state in the sector of the fock space with $Q_1=1$,$Q_2=0$
is given in eq. (2.20). From eq. (A.3) we find, at time $\tau=0$,
$$ A(q,0)=0 \qquad B(p,k,0)=f^*(k)\delta^4(\tilde p-p-k)
   \eqno(A.4)$$
Defining the Laplace transformed coefficients $\tilde A(q,z)$ and
$\tilde B(p,k,z)$ as in eq. (2.23), we use eq. (2.24) and the fact that
in eq. (A.4) $A(q,0)=0$ to obtain
$$\eqalign{\tilde A(q,z)&={i\over h(q,z)} \int d^4k f(k) {{B(q-k,k,0)}\over
  z-{{(q-k)^2}\over 2M_N}-{{k^2}\over 2M_\theta}} \cr
  \tilde B(p,k,z)&=\left(z-{{p^2}\over 2M_N}-{{k^2}\over 2M_\Theta}\right)
  ^{-1}(iB(p,k,0)+f^*(k)\tilde A(p+k,z)) \cr}
  \eqno(A.5) $$
Where $h(q,z)$ is defined in eq. (2.26). Inserting in (A.5) the initial
condition for $B(p,k,0)$ from eq. (A.4) we get
$$\eqalign{\tilde A(q,z)&=i\delta^4(\tilde p-q)\left(
  {{z-\tilde p^2/2M_V}\over h(q,z)}-1\right) \cr
  \tilde B(p,k,z)&=\left(z-{{p^2}\over 2M_N}-{{k^2}\over 2M_\theta}\right)
  ^{-1} if^*(k)\delta^4(\tilde p-p-k){{z-\tilde p^2/2M_V}
  \over h(\tilde p,z)} \cr}
  \eqno(A.6) $$
\par Performing the Laplace transform of eq. (2.20) implied by eq. (A.2), we
use the coefficients from eq. (A.6) and evaluate the resulting expression at
the point $z=\omega_V(\tilde p)-i\epsilon=\tilde p^2/2M_V-i\epsilon$. This
procedure give the simple answer
$$ \lim_{\epsilon\to 0} \int_0^{-\infty}
   d\tau U(\tau)Ve^{i(\omega_V(\tilde p)-i\epsilon)}
   b^{\dagger}(\tilde p) \vert 0 \rangle
   = -ib^\dagger(\tilde p)\vert 0\rangle=-i\vert V(\tilde p) \rangle
   \eqno(A.7) $$
and this implies the desired result.
\bigskip
{\bf Appendix B.}
\smallskip
\par We prove here that the vector valued function $\vert n \rangle_{\sigma,
P}$ is necessarily of the form
$$\vert n \rangle_{\sigma,P} = g_P(\sigma)\vert n \rangle_P
  \eqno(B.1)$$
where $\vert n \rangle_P$ is a fixed (for a given value of $P$) vector in
the auxiliary Hilbert space $H$ of the Lax-Phillips representation of the
relativistic Lee-Friedrichs model. 
\par We start with the observation made at the begining of section 4 (see Eq.
(4.1) and the discussion following it) that the operator valued function
$P_{n,P}(\sigma)$, defined in eq. (3.14), is a projection operator for each
value of $\sigma$

$$P_{n,P}(\sigma) P_{n,P} (\sigma) = P_{n,P}(\sigma)
  \eqno(B.2)$$
It is, therefore, a bounded positive opertor on the real $\sigma$ axis
(indeed, for each $\sigma$, the eigenvalue of $P_{n,P}(\sigma)$ is 1 and the
eigenvector is $\vert n \rangle_{\sigma,P}$ ).
\par In order to proceed we need several definitions and results from
the theory of operator valued functions. We denote the upper half plane of
the complex $\sigma$ plane by $\Pi$. If
$b$ is some separable Hilbert space, we denote by ${\cal B}(b)$ the set of
bounded linear operators on $b$. We define the following sets of
${\cal B}(b)$ valued functions$^{18}$
\smallskip
{\bf Definition A}:

\item {(i)}   A holomorphic ${\cal B}(b)$ valued function $f(\sigma)$ on
      $\Pi$ is of bounded type on $\Pi$ if \break $log^+\vert f(\sigma)
      \vert_{{\cal B}(b)}$
      has a harmonic majorant on $\Pi$. The class of all such functions is
      denoted $N_{{\cal B}(b)}(\Pi)$.

\item {(ii)}  If $\phi$ is any strongly convex function, then by
      ${\cal H}_{\phi,
      {\cal B}(b)}(\Pi)$ we mean the class of all holomorphic ${\cal B}(b)$
      valued functions $f(\sigma)$ on $\Pi$ such that $\phi
      (log^+\vert f(z)\vert_{{\cal B}(b)})$ has a harmonic majorant on $\Pi$.

\item {(iii)} We define $N^+_{{\cal B}(b)}(\Pi) = \bigcup
      {\cal H}_{\phi,{\cal B}(b)}(\Pi)$, where the union is over all strongly
      convex functions $\phi$.

\item {(iv)}  By $H^\infty_{{\cal B}(b)}(\Pi)$ we mean the set of all bounded
      holomorphic ${\cal B}(b)$ valued functions on $\Pi$.

Here $log^+t={\rm max}(log t,0)$ for $t>0$ and $log0=-\infty$. The sets
$N_{{\cal B}(b)}$ and $N_{{\cal B}(b)}^+$ are called Nevanlinna
classes and ${\cal H}_{\phi,{\cal B}(b)}(\Pi)$ is a Hardy-Orlicz class.
\par We will need the following theorems and definitions:
\smallskip
{\bf Theorem A}: we have
$$ H^\infty_{{\cal B}(b)}(\Pi) \subseteq {\cal H}_{\phi,{\cal B}(b)}(\Pi)
   \subseteq N^+_{{\cal B}(b)}(\Pi) \subseteq N_{{\cal B}(b)(\Pi)} $$
\smallskip
{\bf Definition B}: Let $u,v$ be nonzero scalar valued functions in $N^+(R)$
($N^+(R)$ is the boundary function for a scalar Nevannlina class function).
A ${\cal B}(b)$-valued funcution $F$ on $R$ is of class ${\cal M}(u_i,v_i)$
if $uF,vF^* \in N^+_{{\cal B}(b)}(R)$.
\smallskip
{\bf Definition C}: If $A \in H^\infty_{{\cal B}(b)}(\Pi)$ then:

\item {(i)}  A is an {\it inner function} if the operator
$$ T(A) \colon f \to Af, \qquad f \in H^2_b(\Pi)$$
is a {\it partial isometry} on $H^2_b(\Pi)$;
\item {(ii)} A is an {\it outer function} if
$$ \bigcup \{Af \colon f \in H^2_b(\Pi) \}=H^2_M(\Pi)$$
for some subspace $M$ of $b$.
\smallskip
\par The main theorem which we will apply here is the following:
\smallskip
{\bf Theorem B}: Let $v$ be any nonzero scalar function in $N^+(R)$. If F is
any nonnegative ${\cal B}(b)$-valued function of class ${\cal M}(v,v)$ on $R$
then
$$F=G^*G$$
on $R$, where $G$ is an outer function of class ${\cal M}(1,v)$ on $R$.
The factorization of $F$ is essentially uniqe.
\smallskip
\par Since $P_{n,P}(\sigma)$ is a bounded operator then, from definition
A(iv) and theorem A we see that
$$ P_{n,P}(\sigma) \in N^+_{{\cal B}(H)}(\Pi). $$
where $H$ is the auxiliary Hilbert space of the Lax-Phillips represention of
the relativistic Lee-Friedrichs model, defined by the variables $\gamma$ in
eq. (3.3) (or eq. (3.5),(3.6) ). Furthermore, the projection operator
$P_{n,P}(\sigma)$ satisfies $(P_{n,P}(\sigma))^* = P_{n,P}(\sigma)$ and, from
definition C we immediately have
$$ P_{n,P}(\sigma) \in {\cal M}(1,1)$$
We can apply theorem B with the result that there is a unique decomposition of
$ P_{n,P}(\sigma)$ 
$$ P_{n,P}(\sigma) = G^*G=(P_{n,P}(\sigma))^* P_{n,P}(\sigma) =
   P_{n,P}(\sigma) P_{n,P}(\sigma) $$
and that $G=P_{n,P}(\sigma)$ is an {\it outer function}. Denote by $P_{n,p}$
the operator on $H^2_H(\Pi)$ for which the realization as an operator valued
function is $P_{n,p}(\sigma)$. From definition C(ii) we therefore have
$$ \left\{ \bigcup P_{n,P}f \colon f \in H^2_H(\Pi) \right\} = H^2_M(\Pi) $$
where $M$ is a subspace of the auxiliary
Hilbert space $H$. If $\vert f \rangle_\sigma$ is the vector valued function
corresponding to some $f \in H^2_H(\Pi)$ then we can write this explicitly as

$$ \Big\{ \bigcup {{\vert n \rangle_{\sigma,P} \langle n \vert_{\sigma,P} }
   \over {{}_{\sigma,P}}\langle n \vert n \rangle_{\sigma,P} } 
   \vert f \rangle_{\sigma} \colon \vert f \rangle_{\sigma}=f
   \in H^2_H(\Pi) \Big\} = H^2_M(\Pi) $$

\par Since $P_{n,P}(\sigma)$ is a projection operator for each $\sigma$ then its
range for each $\sigma$ is a vector proportionl to $\vert n \rangle_{\sigma,
P}$. Define

$$ m=\Big\{ \vert v \rangle \colon \vert v \rangle \in b \ and \ \vert
   v \rangle = \vert n \rangle_{\sigma,P} \  for\  some\ \sigma
   \in \Pi \Big\}$$

Denote by $M$ the subspace of $b$ spanned by the vectors in m. Clearly
the range of $P_{n,P}$ lies in $M$ and there is no smaller subspace of
$b$ that contains the range of $P_{n,P}$. Therefore we have
                                                                                                                      
$$ \left\{ \bigcup P_{n,P}f \colon f \in H^2_H(\Pi) \right\}
   \subseteq H^2_M(\Pi) $$

If the dimension of $M$ is greater then one, a function
$f \in H^2_M(\Pi)$ can always be found that is not in
$\left\{ \bigcup P_{n,P}f \colon f \in H^2_H(\Pi) \right\}$
(for example, take a specific value $\sigma=\sigma_0$ and define
,for some Hardy class function $g(\sigma)$,
$j(\sigma)=g(\sigma)\vert n \rangle_{\sigma_0,P} $; clearly
$j \in H^2_M(\Pi)$ but $j \not\in \left\{ \bigcup P_{n,P}f \colon f \in
H^2_H(\Pi) \right\}$ ). It is therefore true that

$$ \left\{ \bigcup P_{n,P}f \colon f \in H^2_H(\Pi) \right\}
   \subset H^2_M(\Pi), \qquad \ Dim\ M > 1. $$

Therefore $P_{n,P}$ cannot be an outer function unless $Dim\ M = 1 $.
If $Dim\ M=1$ we have necessarily that
$$\vert n \rangle_{\sigma,P} = g_P(\sigma)\vert n \rangle_P.$$
\bigskip
{\bf Appendix C. }
\smallskip
We prove that in the case of a rational $S$-matrix of the form

$$ S_P(\sigma)=1_H-\vert n \rangle_P{{}_P}\langle n \vert + {{\sigma -
   \overline \mu_P} \over \sigma - \mu_P}
   \vert n \rangle_P{{}_P}\langle n \vert
   \eqno(C.1)$$
the analytic continuation to the complex $\sigma$ plane of the function
defined on the real $\sigma$ axis by

$$ J_P(\sigma)\equiv{{}_{\sigma,P}}\langle n \vert n \rangle_{\sigma,P}=\vert
   g_P(\sigma)\vert^2
   \eqno(C.2) $$
is a constant. 
\par We have proved in section 4 that the Lax-Phillips $S$-matrix in the
relativistic Lee-Friedrichs model has the general form (see Eq. (4.5))

$$ S_P(\sigma)=1_H-\vert n \rangle_P{{}_P}\langle n \vert + 
   {{h(P,\sigma-i\epsilon)} \over h(P,\sigma+i\epsilon)}
   \vert n \rangle_P{{}_P}\langle n \vert
   \eqno(C.3)$$
The function $h(P, \sigma-i\epsilon)/h(P, \sigma+i\epsilon)$ should
then be a
 scalar inner
function. From Eq. (3.12),(4.3) and (C.2) we see that one can write
this function as

$$ {{h(P,\sigma-i\epsilon)} \over h(P,\sigma+i\epsilon)}=
   {{h(P,\sigma+i\epsilon)-2\pi i J_P(\sigma)}
   \over h(P,\sigma+i\epsilon)}=1-2\pi i 
   {{ J_P(\sigma)} \over h(P,\sigma+i\epsilon)}.
   \eqno(C.4)$$
We define

$$ W_P(\sigma)\equiv {\rm Re} \,h(P,\sigma \pm i\epsilon)=
   \sigma-{{P^2}\over 2M_V}-{\cal P}\int d^4k { {\vert f(k)\vert^2} \over
   \sigma - {{(P-k)^2}\over 2M_N} -{{k^2}\over 2M_\theta}}
   \eqno(C.5) $$
Where ${\cal P}$ is the symbol for the principle part of the integral.
Using this definition, the functions $h(P,\sigma \pm i\epsilon)$ 
can be written as

$$ h(P,\sigma \pm i\epsilon)= W_P(\sigma) \pm i\pi J_P(\sigma)
   \eqno(C.6) $$
We have
$$ {{h(P,\sigma-i\epsilon)} \over h(P,\sigma+i\epsilon)}=
   {{W_P(\sigma)-i\pi J_P(\sigma)}\over W_P(\sigma)+i\pi J_P(\sigma)}
   \eqno(C.7)$$
\par In the upper half of the complex $\sigma$ plane we
require that $S_P(\sigma)$ satisfies properties $(a),(b)$ and $(c)$
 listed
 in the
introduction ($S_P(\sigma)$ in then an operator inner function). That
property (c) is satisfied one infers from Eq. (C.7). If we require in (C.4)
that the function $J_P(\sigma)/h(P,\sigma)$ is analytic in the upper half
of the complex $\sigma$ plane then the Lax-Phillips $S$-matrix in
Eq. (C.3) satisfies also property (a). To satisfy property (b) we require that
this function is bounded in the upper half plane,

$$\left\vert {{J_P(\sigma)}\over h(P,\sigma)}\right\vert
  \leq M_1, \qquad ({\rm Im}\sigma>0)
  \eqno(C.8)$$
for some positive bound $M_1$.
\par The adjoint of the $S$-matrix $S^\dagger_P(\sigma)$ given
 ($\sigma$ real) by

$$ S^\dagger_P(\sigma)=1_H-\vert n \rangle_P{{}_P}\langle n \vert + 
   {{h(P,\sigma+i\epsilon)} \over h(P,\sigma-i\epsilon)}
   \vert n \rangle_P{{}_P}\langle n \vert
   \eqno(C.9)$$
is a map from the {\it outgoing} spectral representation to the {\it incoming}
spectral representation. It has the properties $(a'),(b')$ and $(c')$ obtained
from $(a),(b)$ and $(c)$ by replacing everywhere ${\rm Im}\sigma>0$ by
${\rm Im}\sigma<0$ (the Fourier transform of
this operator valued function generates right translation invariant subspaces
of $L^2(0,\infty;H)$). We see from Eq. (C.7) and (C.9) that property $(c')$
is satisfied. Furthermore, we have

$$ {{h(P,\sigma+i\epsilon)} \over h(P,\sigma-i\epsilon)}=
   {{h(P,\sigma-i\epsilon)+2\pi i J_P(\sigma)}
   \over h(P,\sigma-i\epsilon)}=1+2\pi i 
   {{ J_P(\sigma)} \over h(P,\sigma-i\epsilon)}
   \eqno(C.10)$$
In order to satisfy property (a') we require that $J_P(\sigma)/h(P,\sigma)$ is
analytic in the lower half of the complex $\sigma$ plane. Property (b') is
satisfied if this function is also bounded there, {\it i.e.} if

$$\left\vert {{J_P(\sigma)}\over h(P,\sigma)}\right\vert
  \leq M_2 \qquad {\rm Im}\sigma<0
  \eqno(C.11)$$
for some positive bound $M_2$.
\par The function $h(P,\sigma)$, as explicitly defined in $(2.26)$
 (these are actually two diffent functions
in the lower and upper half plane; see for example Eq. (3.12)), has no poles
either in the upper half plane or in the lower half plane. By the
requirements of boundedness Eq. (C.8) and (C.11)
we have that $J_P(\sigma)$ has no poles in the complex $\sigma$ plane. In
this case $J_P(\sigma)$ is an {\it entire function}.
\par In the case of the rational Lax-Phillips $S$-matrix of Eq. (C.1) we have

$$ {{h(P,\sigma-i\epsilon)} \over h(P,\sigma+i\epsilon)}=
   {{\sigma-\overline \mu_P}\over \sigma-\mu_P}
   \eqno(C.12)$$
This immediately implies the following condition (here $\sigma$ is real)

$${\rm Im}\left( (\sigma-\mu_P)h(P,\sigma-i\epsilon) \right)=0
  \eqno(C.13)$$
and, using Eq. (C.6), we have

$$ -\pi (\sigma-{\rm Re}\mu_P)J_P(\sigma)={\rm Im}\mu_PW_P(\sigma)
   ={\rm Im}\mu_P(h(P,\sigma-i\epsilon)+i\pi J_P(\sigma))$$
this can be written in a more compact form

$$ -\pi(\sigma-\mu_P)J_P(\sigma)=h(P,\sigma-i\epsilon){\rm Im}\mu_P
   \eqno(C.14)$$
This relation can be analytically continued to the lower half plane. In the
limit $\sigma \to \infty$ (in the lower half plane) we find

$$ \lim_{\sigma \to \infty} J_P(\sigma) = -{1\over\pi}{\rm Im}\mu(P)
   \qquad {\rm Im}\sigma<0
   \eqno(C.15)$$

The complex conjugate of (C.14) can be analyticaly continued to the upper
half plane and find again that

$$ \lim_{\sigma \to \infty} J_P(\sigma) = -{1\over\pi}{\rm Im}\mu(P)
   \qquad {\rm Im}\sigma>0
   \eqno(C.16)$$

Since $J_P(\sigma)$ in an entire function, Eq. (C.15) and (C.16) imply that
it is a constant
$$J_P(\sigma)=\vert g_P(\sigma)\vert^2= -{1\over\pi}{\rm Im}\mu_P
  \eqno(C.17)$$
This result does not imply that the numerator of $(5.12)$ vanishes.
We have shown that in the rational case, the absolute value of
$g_P(\sigma)$ is constant, but it may have a nontrivial phase.
\bigskip
\noindent
{\bf References}
\frenchspacing
\item{1.}P.D. Lax and R.S. Phillips, {\it Scattering Theory}, Academic
 Press, New York (1967).
\item{2.} C. Flesia and C. Piron, Helv. Phys. Acta {\bf 57}, 697
(1984).
\item{3.} L.P. Horwitz and C. Piron, Helv. Phys. Acta {\bf 66}, 694
(1993).
\item{4.} E. Eisenberg and L.P. Horwitz, in {\it Advances in Chemical
Physics}, vol. XCIX,  ed. I. Prigogine and S. Rice, Wiley, New York
(1997), p. 245.
\item{5.} Y. Strauss, L.P. Horwitz and E. Eisenberg, hep-th/9709036,
submitted for publication.
\item{6.} C. Piron, {\it Foundations of Quantum Physics},
Benjamin/Cummings, Reading (1976).
\item{7.} V.F. Weisskopf and E.P. Wigner, Z.f. Phys. {\bf 63}, 54
(1930); {\bf 65}, 18 (1930).
\item{8.} L.P. Horwitz, J.P. Marchand and J. LaVita,
Jour. Math. Phys. {\bf 12}, 2537 (1971); D. Williams,
Comm. Math. Phys. {\bf 21}, 314 (1971).
\item{9} L.P. Horwitz and J.-P. Marchand, Helv. Phys. Acta {\bf 42}
1039 (1969). 
\item{10.} L.P. Horwitz and J.-P. Marchand, Rocky Mountain Jour. of
Math. {\bf 1}, 225 (1971).
\item{11.} B. Winstein, {\it et al,\ Results from the Neutral Kaon
Program at Fermilab's Meson Center Beamline, 1985-1997\/},
FERMILAB-Pub-97/087-E, published on behalf of the E731, E773 and E799
Collaborations, Fermi National Accelerator Laboratory, P.O. Box 500,
Batavia, Illinois 60510. 
\item{12.} T.D. Lee, R. Oehme and C.N. Yang, Phys. Rev. {\bf 106} ,
340 (1957).
\item{13.} T.T. Wu and C.N. Yang, Phys. Rev. Lett. {\bf 13}, 380
(1964).
\item{14.} L.P. Horwitz and L. Mizrachi, Nuovo Cimento {\bf 21A}, 625
(1974); E. Cohen and L.P. Horwitz, hep-th/9808030; hep-ph/9811332,
 submitted for publication.
\item{15.}  W. Baumgartel, Math. Nachr. {\bf 69}, 107 (1975);
 L.P. Horwitz and I.M. Sigal, Helv. Phys. Acta
 {\bf 51}, 685 (1978); G. Parravicini, V. Gorini
 and E.C.G. Sudarshan, J. Math. Phys. {\bf 21}, 2208
 (1980); A. Bohm, {\it Quantum Mechanics: Foundations
 and Applications\/,} Springer, Berlin (1986); A. Bohm,  M. Gadella
and
 G.B. Mainland, Am. J. Phys. {\bf 57}, 1105 (1989); T. Bailey and
 W.C. Schieve, Nuovo Cimento {\bf 47A}, 231 (1978).
\item{16.} I.P. Cornfield, S.V. Formin and Ya. G. Sinai,
{\it Ergodic Theory}, Springer, Berlin (1982).
\item{17.} Y. Four\`es and I.E. Segal, Trans. Am. Math. Soc. {\bf 78},
385 (1955).
\item{18.} M. Rosenblum and J. Rovnyak, {\it Hardya Classes and
Operator Theory}, Oxford University Press, New York (1985).
\item{19.} L.P. Horwitz, Found. of Phys. {\bf 25}, 39 (1995). See
also, D. Cocolicchio, Phys. Rev. {\bf D57}, 7251 (1998).
\item{20.} E.C.G. Stueckelberg, Helv. Phys. Acta {\bf 14}, 322, 588
(1941); J. Schwinger, Phys. Rev. {\bf 82}, 664 (1951); R.P. Feynman,
        Rev. Mod. Phys. {\bf 20}, 367 (1948) and Phys. Rev. {\bf 80},
        440 (1950); L.P. Horwitz and C. Piron, Helv. Phys. Acta {\bf
        46}, 316 (1973); R. Fanchi, Phys. Rev {\bf D20},3108 (1979);
        A. Kyprianides, Phys. Rep. {\bf 155}, 1 (1986) (and references
 therein).
\item{21.} I. Antoniou, M. Gadella, I. Prigogine and P.P. Pronko,
        Jour. Math. Phys. {\bf 39}, 2995 (1998).
\item{22.} N. Shnerb and L.P. Horwitz, Phys. Rev. {\bf A48}, 4068
(1993); L.P. Horwitz and N. Shnerb, Found. of Phys. {\bf 28}, 1509 (1998).
\item{23.} For example, J.R. Taylor, {\it Scattering Theory},
 John Wiley and Sons,
 N.Y. (1972); R.J. Newton, {\it Scattering Theory of Particles
 and Waves}, McGraw Hill, N.Y. (1976).
\item{24.} K. Hoffman, {\it Banach Spaces of Analytic Functions}
 Prentice Hall, Englewood Cliffs, N.J. (1962).
\item{25.} T.D. Lee, Phys. Rev. {\bf 95}, 1329 (1954);
K.O. Friedrichs, Comm. Pure and Appl. Math. {\bf 1}, 361 (1950).  This
result was originally conjectured by C. Piron (personal
communication).
\item{26.} Y. Strauss and L.P. Horwitz, in preparation.
\item{27.} H. Lehmann, K. Symanzik and W. Zimmerman, Nuovo Cim. {\bf
1}, 1425 (1955). 
        
\vfill
\eject
\end
\bye